\documentclass[11pt]{article}
\usepackage[left=3cm, right=3cm]{geometry}
\usepackage{authblk}
\usepackage{algorithm}
\usepackage{algpseudocode}
\usepackage[numbers]{natbib}
\usepackage[english]{babel}
\usepackage[utf8]{inputenc}
\usepackage[T1]{fontenc}
\usepackage{amsmath}
\usepackage{amssymb}
\usepackage{graphicx}
\usepackage[english]{babel}
\usepackage{csquotes}
\usepackage{xcolor}
\usepackage{natbib}
\usepackage{subfigure}
\usepackage{float}

\usepackage[colorlinks=true,linkcolor=blue]{hyperref}
\hypersetup{urlcolor=blue, citecolor=blue, colorlinks=true, linkcolor=blue}

\newcommand{\round}[1]{\left( #1 \right)}

\newcommand{\curly}[1]{\left\{ #1 \right\}}

\newcommand{\Poi}[1]{\text{Poi}\round{#1}}
\newcommand{\NegBin}[1]{\text{NegBin}\round{#1}}

\author[1]{Tobia Filosi}
\author[2,3]{Emiliano Ceccarelli\thanks{Corresponding author: Emiliano Ceccarelli, email: nome.cognome@uniroma1.it}}
\author[4]{Emilio Porcu}
\author[3]{Elena Del Sordo}
\author[5]{Libia Lara-Carri{\'o}n}
\author[6]{Giuseppe Iuppa}
\author[7]{Francesca Puoti}
\author[7]{Silvia Trapani}
\author[7]{Silvia Testa}
\author[2]{Giovanna Jona Lasinio}

\affil[1]{Department of Mathematics, University of Trento.}
\affil[2]{Department of Statistical Sciences, Sapienza, University of Rome.}
\affil[3]{Statistical Services, Italian National Institute of Health, Rome, Italy}
\affil[4]{Department of Mathematics and BTC at Khalifa University.}
\affil[5]{College of Medicine and Health Sciences, Khalifa University.}
\affil[6]{Cleveland Clinic at Abu Dhabi}
\affil[7]{Italian National Transplant Center, Italian National Institute of Health, Rome, Italy}

\title{Scalable model selection for count time series with structural breaks: application to solid-organ transplantation during and after COVID-19 in the USA and Italy}

\begin{document}
\maketitle

\begin{abstract}
\textbf{Background:} Weekly healthcare activity data are typically non-negative counts with temporal dependence and occasional system-wide disruptions, settings in which Gaussian time-series models may be inadequate. Solid organ transplant (SOT) activity provides a representative case study of a count process affected by a large external shock.

\textbf{Methods:} We analyse weekly SOT counts in the USA and Italy from 2014 to October 2024, stratified by donor type (deceased vs living) and organ (kidney and liver). We fit Poisson and negative-binomial count time-series models incorporating short-term dynamics, calendar effects (holiday weeks), and pre-specified pandemic-period level and/or slope indicators. Candidate specifications are screened within a pre-defined portfolio and selected using BIC within each training window. Forecasting performance is evaluated with an expanding-window design at horizons $h\in\{4,8,12\}$ weeks. Alongside RMSE, we report empirical coverage of nominal $95\%$ predictive intervals and interval widths to summarise calibration and forecast uncertainty.

\textbf{Results:} Across strata, selected models capture substantial pandemic-period deviations and varying post-period trajectories. Deceased-donor series are broadly consistent with a return towards pre-pandemic baselines in both countries, whereas the US living-donor series shows a more gradual convergence in this application. Within the explored model class and validation protocol, auxiliary covariates representing COVID burden and mortality add limited incremental predictive contribution beyond autoregressive and calendar components.

\textbf{Conclusions:} Our analysis shows that donation time series represent an unconditional phenomenon, with auxiliary variables having a statistically negligible impact on donations, thus allowing a focus on more practical aspects related to ongoing challenges in the post-pandemic era, such as hospital overloads and changes in public perception.
\end{abstract}

\section{Background}
This work is motivated by a common problem in medical research methodology: weekly healthcare activity data are non-negative counts, often overdispersed, and can be affected by abrupt system-wide shocks (e.g., the SARS-CoV-2 pandemic). Standard Gaussian time-series models are typically inadequate in this setting. We therefore propose a reproducible, calibration-first protocol for selecting and validating count time-series models for weekly transplant activity, and we illustrate it using solid organ transplant (SOT) counts in the USA and Italy.

\subsection{Context}
    Solid Organ Transplants (SOT) have registered an increasing trend in the pre-pandemic era. Several countries have celebrated the increasing acceptance of organ transplants by the population of both deceased and living donors. SARS-COV2 did not make an exception with SOT and caused a relevant shock that reflected in an appalling crash in the number of donations per week. This is not a mystery, and the crash caught the attention of disciplines as diverse as medicine, biology, statistics, and AI.  \\
    With the advent of the pandemic, the American Society of Transplant Surgeons (ASTS) recommended suspending all living donations unless strictly necessary \citep{ali2021safe}. Despite a major effort by Centres for Medicare and Medicaid services to avoid postponing organ transplants, USA experienced a major crash in donations. The crash happened despite SOT being classified as a {\em tier} $3b$. In turn, ASTS classified as {\em safe} to use organs from donors who were low-risk or COVID-19 negative, as well as those who have recovered from COVID-19 disease more than 28 days before donation \citep{gatti2022clinical}. During the initial months of the COVID-19 outbreak, all the efforts of the Italian National Transplant Center (CNT), the competent authority in the donation and transplantation field, were focused on the preservation of donation and transplantation activities, as urgent and life-saving procedures calling for continuity. Regulatory measures were issued to safely continue this activity in Italy, by routinely testing donors and recipients on the waiting list for SARS-CoV-2 and creating COVID-free pathways inside the transplant centers \citep{trapani2021incidence}.
    
    While the attention from the scientific community on such a problem is unquestionable, the success of scientific initiatives related to the problem is definitely scarce. Existing work has not fully clarified why the crash occurred - apart from the pandemic, was there something more we should be worried about? Further, the current literature provides limited evidence on whether pre-pandemic trends will persist and, if so, at what pace. \\
    {This paper focuses on model choice and distributional predictive validation in the presence of sudden shocks, aiming to assess whether and how the pre-pandemic increasing trends may resume. Our effort proves that the answers are far from being trivial. We embrace a comprehensive portfolio of covariates and challenge a huge model-selection problem to shade some lights on the plethora of questions coming from practitioners all over the world. While such a selection might be taken through deterministic algorithmic procedures, we are keen on keeping statistical accuracy to safely understand the origins of trends and their variability. This entails not only preserving accuracy in model selection, but also in estimation performance and prediction accuracy.} 

\subsection{The state of the art} 

    The standing literature has shown consistent interest in the time series of SOT's counts. 
    \cite{Ibrahim:2021} focus on the ethical aspects of organ and tissue donations during the pandemic, highlighting the trade-off between infection risk and saving lives. As virus knowledge advanced, ethical dilemmas have been largely addressed, aiding healthcare recovery.\par
    Several studies document the pandemic's dramatic impact on SOT. \cite{AUBERT:2021} analysed data from 22 countries, noting significant variations in transplant activity across different types. Kidney transplants were most affected, followed by lung, liver, and heart transplants, with universal decreases but varying responses among countries. However, the simplistic statistical methods used in these studies do not allow for thorough inspections of historical dynamics or reliable future predictions.\par
    \citet{Nimmo:2022} examined spatial heterogeneity in the impact of COVID-19 on organ transplants, finding significant reductions in some countries despite low COVID-19 death rates and moderate declines in others despite higher death rates. Deceased donor transplantation showed reductions across all donation stages, especially early in the pandemic. Living donor transplantation saw a more pronounced reduction, with global declines of 40\% in kidney transplants and 33\% in liver transplants in 2020 \cite{AUBERT:2021}. Similar findings are reported in Indian cohorts by \cite{KUTE:2022}. However, these studies often lack rigorous statistical methods, affecting their accuracy and reliability.\par
    More robust statistical approaches have been used in some studies. \citet{SUAREZ:2022}, focusing on USA data, developed an autoregressive-integrated-moving-average (ARIMA) model using monthly data from 1990 to 2019 to forecast expected transplantation rates for 2020 in a hypothetical scenario without the pandemic. They identified a significant transplantation deficit during the pandemic, particularly affecting kidney transplantation and waitlist registrations.\par
    {A meta-analysis of the clinical outcome in SOT recipients affected by COVID-19 compared to the general population proved \citep{coll2021covid} that SOT recipients affected by COVID-19 were not associated with an increased risk of mortality compared with the general population. Transplant recipients may be at a higher risk of infection by SARS-CoV-2 due to the baseline use of immunosuppression, underlying comorbidities, and frequent contact with the healthcare system. A study done in the US \citep{coll2021covid} showed that those with a kidney transplant have the highest risk of acquiring COVID-19, and comorbidities are common in all SOT patients. Yet, they are even more common in those testing positive for COVID-19. SOT patients who test positive for COVID-19 are at high risk for hospitalization, major adverse renal or cardiac events, and acute kidney injury. According to \citep{vinson2021covid}, the decrease in the SOT during COVID was due to 59$\%$ of programs pausing living donor evaluation. UNOS database indicates that the centre-wide acceptance rate of organs from COVID+ donors is increasing but is variable with current rates of 25$\%$ for heart, 41$\%$ for kidney, and 51$\%$ for liver transplants. Antiviral drugs such as nirmatrelvir/ritonavir can be a promising therapeutic option for SOT recipients with COVID-19, and the proportion of SARS-CoV-2 positive donor referrals who then became donors has increased.}\par
    Similarly, \cite{Porcuet:2022} used an ARIMA model to analyse weekly SOT data from the USA, incorporating COVID-19 cases as a covariate. Their model accurately estimated the correlation structure and predicted future donor activity for deceased and living donors. To our knowledge, there are no published time-series analyses quantifying the impact of COVID-19 on transplant activity in Italy; available contributions are mainly clinical or organizational reports, such as the early Italian experience described by \cite{angelico2020covid}, which noted an initial reduction in deceased donor procurement during the first weeks of the outbreak.
    
    Thus, the existing literature highlights COVID-19's significant impacts on SOT, which are not fully supported by rigorous statistical protocols. Our approach aims to address these gaps by {providing} more sophisticated {and robust} statistical models to {have accurate modelling, estimation, and prediction of the relevant} trends in organ donations. There is a consensus that counts data time series are best modelled using Poisson or negative binomial-based models for count data time series \citep{FOKIANOS:2011, Chistou:2014, liboschik_tscount_2017}. These models provide a robust framework for analysing count data, ensuring accuracy in statistical inference.

    \subsection{Contribution}

    Modelling time series in the presence of sudden shocks is a major challenge. Trends become uninterpretable, and variability explodes. So much so that Gaussian time series models are often inadequate to assess what happened, why it happened, and what it will happen. This translates into a major challenge in terms of model selection and prediction.
     In this study, we jointly analyse transplant dynamics in the USA and Italy, applying the same modelling framework to both national contexts. This allows us to assess common patterns and country-specific divergences in donation trends, particularly in the aftermath of the SARS-COV2 pandemic.
    To embark on such a challenge, we trained more than $100,000$ parametric models associated with time series of counts framework with auxiliary variables to account for potential confounding effects. Not only does our effort entail accuracy, but it also deals with a high-dimensional framework for which there is a notorious trade-off between accuracy and computational scalability. Specifically we develop a a practical, reproducible protocol that integrates candidate model design, information-criterion screening, and forecast calibration checks for count time series with interventions. We model weekly SOT counts using Poisson/negative-binomial count time-series models, allowing for autoregressive dynamics, calendar effects (holiday weeks), and level and/or slope changes associated with the COVID-19 period. The intervention indicators are used as pragmatic break proxies to improve predictive adequacy and summarise regime changes; we do not claim causal attribution.
    \section{Methods}
\subsection{Data availability}
    {The time series considered in this paper contains SOT counts for both deceased and living donors. The time horizon is about a decade, from January 2014 until the week of October 21, 2024. The frequency of observation is weekly. As a consequence, our data set is not necessarily to be considered as a {\em big data}. Yet, the problem considered in this paper requires a {big} computational effort: the combination of covariates with potential model parametrisations provides a massive research portfolio.
    Achieving statistical accuracy is extremely challenging for every problem, requiring a consistent computational effort. This is one of our main goals.
    {In this work, we focus on kidney and liver SOT as the two types of transplants represent the large majority of transplants from deceased (In the US about 80\% of which 55\% kidney and 25\% liver, and in Italy about 88\% of which 51\% kidney and 38\% liver) and living donors (in the US 90\% kidney and between 5 and 9\% liver, and in Italy 92\% kidney and 8\% liver). }\\

    USA data are available thanks to the Organisation for Economic Co-operation and Development (OECD) data repository \cite{deathsdata} and thanks to the World Health Organization (WHO) data dashboard \cite{coviddatawho}.
    Italy's transplant data are available thanks to the National Transplant Centre of the Italian National Health Institute (ISS), additional covariates are available online thanks to the Italian National Statistics Institute (ISTAT) and the Italian Civil protection.\\
    Potentially influential factors include the total number of all-cause deaths (OECD and ISTAT), the estimated number of all-cause deaths (as a baseline), the number of new COVID-19 cases, the number of COVID-19 deaths, the number of non-COVID-19 deaths (WHO and Italian Civil protection), and {{\em time}, which is customarily represented by a progressive number. Data have been cleansed and standardised to make them homogeneous and thus allow for comparison.} %

    \subsection{Count time-series model}
    
    Our modelling effort is based on the Poisson and Negative Binomial distributions. Count data may be distributed as a Negative Binomial if the rate at which events occur is heterogeneous, that is: the counts are characterised by overdispersion compared to the Poisson (as typically happens in the number of deaths time series). The Poisson distribution is nested within the Negative Binomial, in the sense that if no overdispersion/heterogeneity is present, the Negative Binomial distribution converges to the Poisson distribution.

Let $Y_t$ be a discrete count time series (that is: $t\in\curly{1,\dots,T}$ and $Y_t$ be a non-negative integer for all $t$), whose dynamic may depend on a covariate vector $X_t$. It is assumed that $Y_t$, conditionally on the past $\mathcal F_{t-1}$, follows either a Poisson or a negative binomial distribution:
\begin{equation}
    Y_t \big| \mathcal{F}_{t-1}\sim \Poi{\mu_t} \qquad Y_t \big| \mathcal{F}_{t-1}\sim \NegBin{\mu_t, \phi},
\end{equation}
where $\mu_t$ is the conditional mean and $\phi$ is the overdispersion parameter. 

Several models are proposed for the parameter $\mu_t$: the link can be either the identity or the logarithm function, and the covariates can show both an internal or external effect. While the link determines whether the mean $\mu_t$ is considered directly or via its logarithm, the effect of covariates determines whether they are taken into account in the recursive computation of $\nu_t$ or not, as specified below. In the following, $X_t$ is a vector of covariates, $\varphi(\cdot)$ is either $\text{id}(\cdot)$ or $\log(\cdot + 1)$ and it is applied element-wise. In addition, $s$ is a non-negative integer whose role is to lag the effect of the covariates $X$. Finally, $\Delta_t$ is a vector of special covariates, for which taking the log or applying a lag does not make sense, like the dummies and the linear trend.\par
In the following, $C_t$ is the COVID-19 dummy, defined as $0$ for $t\leq (2019,52)$ and $1$ for $t\geq (2020,1)$. In our framework, we model the structural break at $t=(2020, 1)$ through parametric intervention terms driven by $C_t$, allowing for a possible level shift and/or slope change, rather than estimating an unconstrained change-point location.
Similarly, $\eta_{time,t}$ is the slope coefficient, which may vary during the pre and post COVID-19 period. Formally, it can be defined as $\eta_{time,t}:=\eta_{time} + \delta_{slope}\Delta\eta_{time}\cdot C_t$, where $\delta_{slope}$ is a binary variable that allows for choosing whether we allow for different trends pre and post pandemic or not. We now present the algorithmic workflow adopted in our model selection procedure, explicitly listing the candidate specifications for the conditional mean $\mu_t$. Model selection is performed using a pre-specified portfolio and objective criteria, BIC \citep{schwarz1978estimating} screening and rolling-origin forecast evaluation. Nevertheless, coefficient-level uncertainty summaries should be interpreted as conditional on the selected specification: they do not incorporate additional uncertainty due to the model-search step. Accordingly, our main conclusions emphasise out-of-sample predictive performance and predictive interval behavior rather than hypothesis testing. \par
If the link is the \emph{identity} and the covariates are \emph{internal}, the model can be formalised as:
\begin{align*}
    \mu_t &:= \nu_t + \Delta\beta_0\cdot C_t + \eta_{time,t}t\\
    \nu_t &= \beta_0 + \sum_{k=1}^{p} \beta_k Y_{t-i_k} + \sum_{\ell=1}^{q} \alpha_\ell \nu_{t-j_\ell} + \eta^\top \varphi(X_{t-s}),
\end{align*}
whilst, in the case covariates are \emph{external}, the model is
\begin{align*}
    \mu_t &:= \nu_t + \Delta\beta_0\cdot C_t + \eta_{time,t}t+\eta^\top \varphi(X_{t-s})\\
    \nu_t &= \beta_0 + \sum_{k=1}^{p} \beta_k Y_{t-i_k} + \sum_{\ell=1}^{q} \alpha_\ell \nu_{t-j_\ell}.
\end{align*}
Similarly, if the link function is \emph{logarithmic}, then the model becomes
\begin{align*}
    \mu_t &:= \exp(\nu_t+\Delta\beta_0\cdot C_t +\eta_{time,t}t)\\
    \nu_t &= \beta_0 + \sum_{k=1}^{p} \beta_k \log(Y_{t-i_k}+1) + \sum_{\ell=1}^{q} \alpha_\ell \nu_{t-j_\ell} + \eta^\top \varphi(X_{t-s})
\end{align*}
for covariates with an \emph{internal} effect, while it reads
\begin{align*}
    \mu_t &:= \exp\round{\nu_t + \Delta\beta_0\cdot C_t+\eta_{time,t}t+ \eta^\top \varphi(X_{t-s})}\\
    \nu_t &= \beta_0 + \sum_{k=1}^{p} \beta_k \log(Y_{t-i_k}+1) + \sum_{\ell=1}^{q} \alpha_\ell \nu_{t-j_\ell}
\end{align*}
for covariates with an \emph{external} effect. Clearly, the indices $i_1,\dots,i_p$ and $j_1,\dots,j_q$ are positive integers that rule which past observations and means are to be considered. 

For each variable of interest (SOT from deceased and living donors), we explored all the combinations of the following model's characteristics:
\begin{itemize}
    \item distribution: Poisson or negative binomial;
    \item $p$: number of past observations (autoregressive terms): from 0 to 8; 
    \item $q$: number of autoregressive terms for the mean: from 0 to 8;
    \item link function: identity or logarithm;
    \item effects of covariates: external or internal;
    \item $\varphi$: scale of the standard covariates $X$: identity or logarithm; 
    \item $s$: lag of standard covariates $X$: from 0 to 8;
    \item covariates $X$: total number of all-causes deaths, estimated number of all-causes deaths, number of new COVID-19 cases, number of COVID-19 deaths, number of not COVID-19 deaths;
    \item special covariates (dummies): changing point in trend slope ($\delta_{slope}$): this special covariate has always been assumed to have an external effect and no lag. In addition, we have always allowed for a change in the intercept ($\Delta\beta_0$).
\end{itemize}
For simplicity, we chose $i_k:=k$ and $j_\ell:=\ell$ for each $k,\ell$. This means, for instance, that for the model with identity link, $p=2$, $q=3$ and no covariates, the model will read
\begin{align*}
    &Y_t \big | \mathcal{F}_{t-1} \sim \text{Poi}(\nu_t + \delta_j + \eta_{time,j}t),\\
    &\nu_t=\beta_0+\beta_1 Y_{t-1} + \beta_2 Y_{t-2} + \alpha_1 \nu_{t-1} + \alpha_2 \nu_{t-2} + \alpha_3 \nu_{t-3}. 
\end{align*}
Once a model has been estimated, we removed all the indices $i_k$ and $j_\ell$ which showed a non-significant effect. After that, we re-estimated the model with the new parameters. We clarify this procedure with an example. Assume that, after estimating the previous model, the estimates of $\beta_1$, $\alpha_1$ and $\alpha_2$ are statistically non-significant. Then the following model is re-estimated:
\begin{align*}
    &Y_t \big | \mathcal{F}_{t-1} \sim \text{Poi}(\nu_t + \delta_j + \eta_{time,j}t),\\
    &\nu_t=\beta_0 + \beta_2 Y_{t-2} +  \alpha_3 \nu_{t-3}. 
\end{align*}
In this way, we are able to explore also models with non-contiguous indices $i_k$ and $j_\ell$ for past observations and past mean respectively.

Overall, we estimated $108040$ models for total SOT from both deceased and living donors. 2 distribution, 73 combinations of $p$ and $q$, 2 values for the link, 37 covariance settings (9 possible covariates lags $s$, external or internal covariates effect, 2 values for $\varphi$, plus the case without covariates), 5 covariates possibilities, 2 possibilities of COVID-19 effects in the trend (pre-post covid change in trend speed). BIC provided a computationally efficient first-stage screening of a large candidate set \par
Regarding the models for kidney and liver SOT, we decided not to include any covariate, since they have not shown any statistically significant effect for the total number of SOT. As a consequence, for each combination (kidney and liver SOT from both deceased and living donors) we estimated 584 models. 2 distributions, 73 combinations of $p$ and $q$, $2$ values for the link function, $1$ covariance settings (no covariates), 2 combinations of COVID-19 effects (change in the slope and intercept or change in the intercept only). The generic algorithm, including the forecast steps (Section \ref{sec:forecast-eval}), is presented here:

\begin{algorithm}[H]
\caption{Model selection and distributional predictive validation}
\label{alg:workflow_compact}
\begin{algorithmic}[1]
\Require Weekly count time series $T_t$; covariates $X_t$; COVID dummy $C_t$ and time trend $\eta_{time,t}$;
families $\{\text{Poi}, \, \text{NegBin}\}$; links $\{\text{id}(\cdot),\log(\cdot+1)\}$; candidate autoregressive terms $\{\beta_k\}$ and $\{\alpha_\ell\}$;
forecast horizons $\mathcal{H}=\{4,8,12\}$.
\Ensure Selected/pruned model and in-sample and out-of-sample RMSE and empirical coverage.

\State \textbf{Portfolio.} Build the candidate set $\mathcal{M}$ by all combinations of $(\beta_k,\alpha_\ell)$, family, link and subsets of $X_t$, COVID dummy $C_t$ and time trend $\eta_{time,t}$.
\State \textbf{Selection.} Fit all $m\in\mathcal{M}$ and select $m^\star=\arg\min_{m}\mathrm{BIC}(m)$.
\State \textbf{Pruning.} From $m^\star$, drop non-significant $\beta_k$, $\alpha_\ell$ and covariates in $X_t$ (keep $C_t$ and $\eta_{time,t}$) $\Rightarrow$ final model $\tilde m$.
\State \textbf{In-sample.} Using $\tilde m$, compute RMSE and empirical coverage of nominal 95\% confidence intervals.
\State \textbf{Out-of-sample.} For each $h\in\mathcal{H}$ forecast $y_{t_0+h}$, $t_0$ fixed, and compute RMSE$_h$ and coverage$_h$.

\end{algorithmic}
\end{algorithm}

\subsection{Statistical and Machine Learning analysis}
    Several aspects should be considered to provide a comprehensive statistical analysis. Time series of SOT counts have very specific features. As noticed by  \cite{Maruottietal:2022} and \cite{Hilbe_2014}, these data's mean and variance are not independent. Hence, classical statistical procedures do not take into account this case, as mean and variance are typically supposed to be mutually independent. A second relevant aspect is that counts are non-negative and discrete. This fact considerably reduces the number of parametric families of probability distributions that might be considered candidates to represent the underlying stochastic mechanism generating the data. Typical time series models work under the assumption that the support of the time series is the whole real line and are often based on the unplausible assumption of Gaussianity \citep{timeSeries:2019}. 

    While there is an established methodology for time series of counts \citep{FOKIANOS:2011}, we surprisingly found that the relevant literature on SOT has never considered a proper modelling assumption for this kind of time series. \par
    The portfolio becomes rich enough when incorporating covariates into the modelling assumption: we provide a wealth of modelling alternatives, summing up over $100,000$ potential choices, whose detailed count and explanation can be found in the Supplementary material. The software implementation of these methods can be achieved through open-source packages \citep{liboschik_tscount_2017}. We provide a baseline model to estimate the expected number of weekly deaths. This is achieved through a linear mixed model, which follows the same criteria as in \cite{Maruottietal:2022} and \cite{linearmixedModel:2021}. Once a baseline is provided, the subsequent step is to model variability and uncertainty. This is crucial to provide a reasonable performance in terms of both estimation and prediction. 
\subsubsection{Model selection over a massive portfolio}
    Our search for optimality embarks on both model fitting and prediction for both types of donors. We start with a portfolio of potential modelling strategies. Then, a comparison regarding fitting performance is provided through the Bayesian Information Criterion (BIC) \citep{BIC}. As for predictive performance, a classical mean squared error assessment is provided (RMSE, see \cite{gneiting:2007}).  \par
    A deeper inspection of the model's ability to capture the time series dynamics is provided through the \emph{empirical coverage}. For each model, we consider a part of the data set as a training set. Then, for each model, prediction is performed, and the predicted value is compared with the {\em true} value. As a result, one can compute the proportion of observed data that fall within a $95\%$ confidence interval (see Supplementary material Tables \ref{tab:dec} -- \ref{tab:livKITA} and Table \ref{tab:prediction} for the empirical coverage at different time horizons). \par
    The modelling portfolio considered in this paper makes our procedures computationally very expensive.
    We used the available data to quantify the effect of the pandemic on the SOT. \emph{A priori}, all these variables could significantly describe the phenomenon. The all-cause mortality rate could be influential on the number of SOT from deceased donors. Yet, the expected number of deaths (baseline) could be a neater indicator,  {being} less affected by potential mortality peaks due to extreme events. {Further,} it can be available in advance if we are interested in predicting SOT. The number of new COVID cases and deaths may affect the whole healthcare system and, therefore, potentially influence SOT. Finally, time has been introduced to model the linear increase in SOT.
    
    \subsection{Expanding window forecast evaluation}
\label{sec:forecast-eval}
To assess predictive performance under realistic deployment, we use a fixed-origin expanding-window procedure. For each horizon $h \in \mathcal{H} =\{4,8,12\}$, we evaluate forecasts over a test period of $n$ consecutive weeks: at each week $t$ in the test set, the model is fitted on $\{1,\ldots,t-h\}$ (expanding window) and used to predict $Y_t$. $RMSE_h$ and $coverage_h$ are then computed by averaging over the $n$ test weeks. The choice of $\mathcal{H} $was dictated by the fact that longer-range forecasts would be dominated by uncertainty accumulation and are less credible in this setting. Consistently, the models most frequently selected by our procedure exhibit only short autoregressive structures, suggesting limited predictability beyond a few weeks. For each selected model we produce $h$-step-ahead predictive distributions and point forecasts.
We construct nominal 95\% predictive intervals by taking the $a$- and $(1-a)$-quantiles of the (approximated) predictive distribution, with $a=(1-0.95)/2$.

\subsection{Implementation}
Dealing with these computational aspects required the implementation of an iterative procedure in concert with the use of high-performance computers at the University of Trento and the University of Rome Sapienza. {This allowed} to consider all possible combinations of influential factors and {to } account for pre- and- post-COVID changes in trends and relations with auxiliary variables. All analyses were conducted using \texttt{R} \citep[version~4.4.2,][]{Rstudio}. We utilised the \texttt{tscount} packages \citep{liboschik_tscount_2017}, {that is} specifically designed for handling count time series. We parallelize the model-fitting tasks by distributing the elements across all available CPU cores using the \texttt{mclapply} function of the \texttt{parallel} R package \citep{schmidberger2009state}, where each worker independently estimates the model on a distinct candidate specification.

\section{Results}
\subsection*{Explanatory Data Analysis}
    \begin{figure} 
        \centering
        \subfigure[Total SOT from deceased donors.]{
            \includegraphics[width = 0.45\textwidth,height=0.25\textheight]{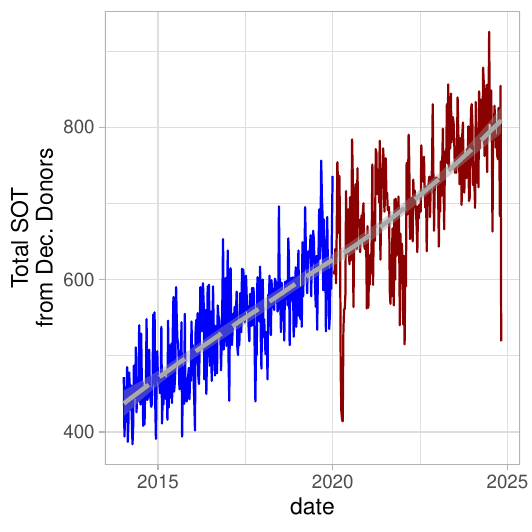}
        }
        \hfill
        \subfigure[Total SOT from living donors.]{
            \includegraphics[width = 0.45\textwidth,height=0.25\textheight]{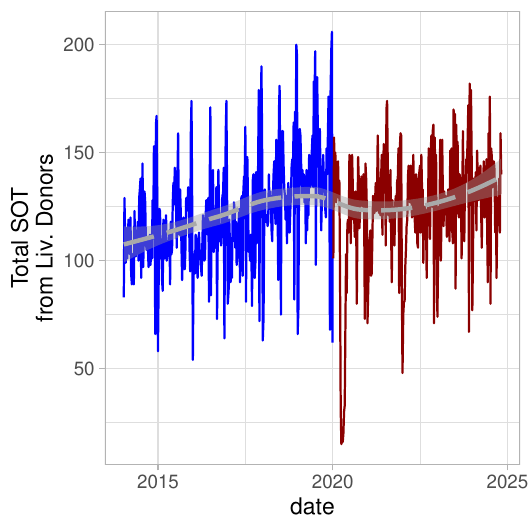}
        }
        \subfigure[Kidney SOT from deceased donors.]{
            \includegraphics[width = 0.45\textwidth,height=0.25\textheight]{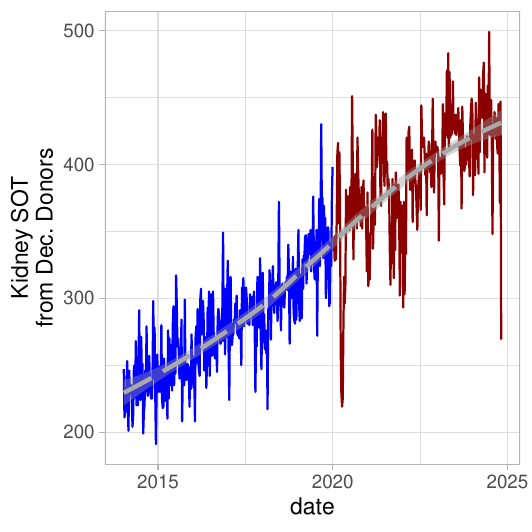}
        }
        \hfill
        \subfigure[Kidney SOT from living donors.]{
            \includegraphics[width = 0.45\textwidth,height=0.25\textheight]{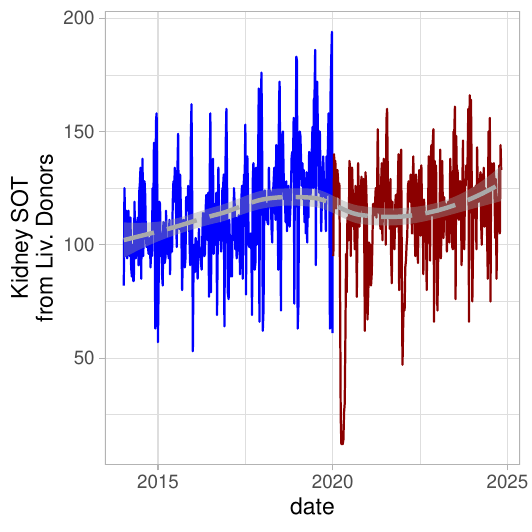}
        }

        \subfigure[Liver SOT from deceased donors.]{
            \includegraphics[width = 0.45\textwidth,height=0.25\textheight]{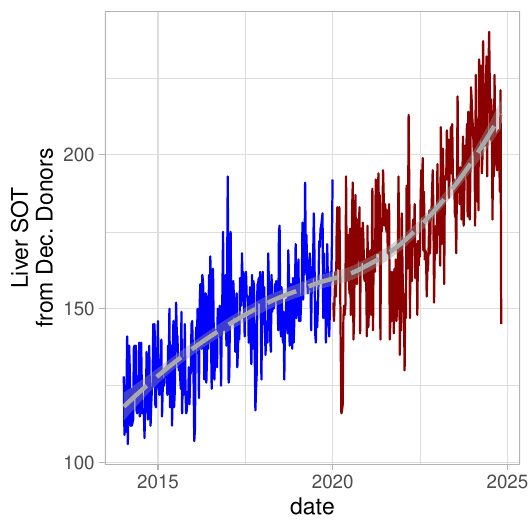}
        }
        \hfill
        \subfigure[Liver SOT from living donors.]{
            \includegraphics[width = 0.45\textwidth,height=0.25\textheight]{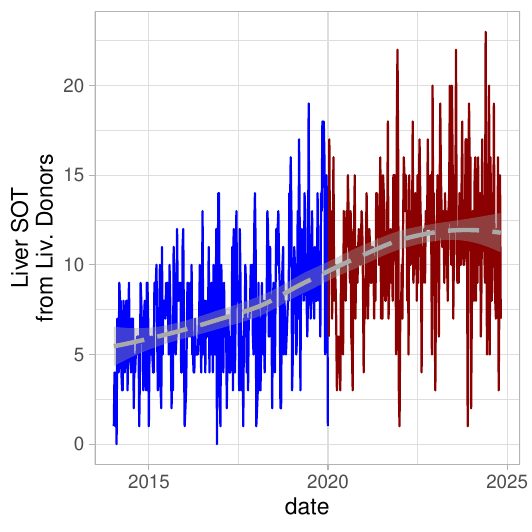}
        }
        \caption{Number of weekly transplants in the USA from deceased ((a),(c),(e)) and living ((b),(d),(f)) donors, comparing pre (blue line) and post (red line) COVID-19 time windows. Trends evaluated as loess curves are grey dashed lines.}
        \label{fig:dec-liv_USA}
    \end{figure}

    \begin{figure} 
        \centering
        \subfigure[Total SOT from deceased donors.]{
            \includegraphics[width = 0.45\textwidth,height=0.25\textheight]{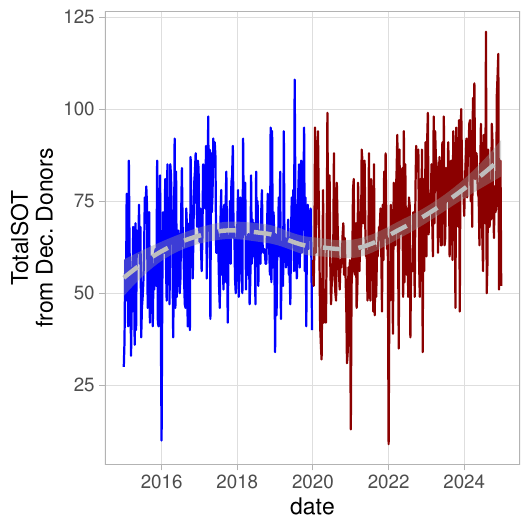}
        }
        \hfill
        \subfigure[Total SOT from living donors.]{
            \includegraphics[width = 0.45\textwidth,height=0.25\textheight]{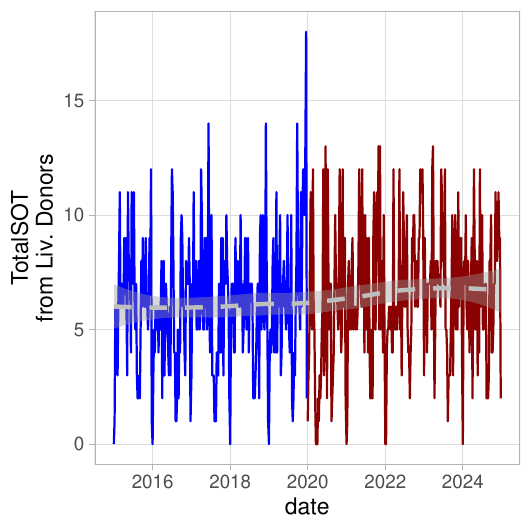}
        }
        \subfigure[Kidney SOT from deceased donors.]{
            \includegraphics[width = 0.45\textwidth,height=0.25\textheight]{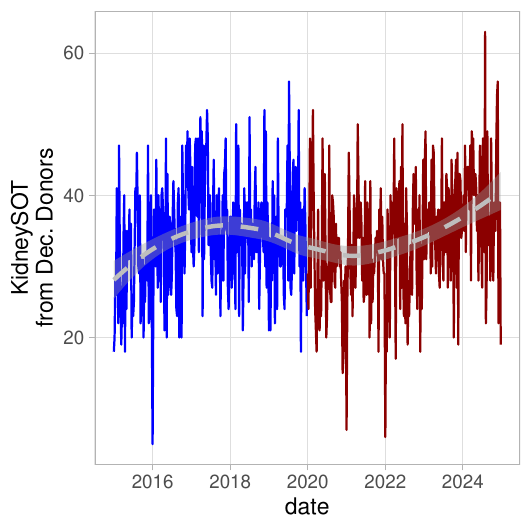}
        }
        \hfill
        \subfigure[Kidney SOT from living donors.]{
            \includegraphics[width = 0.45\textwidth,height=0.25\textheight]{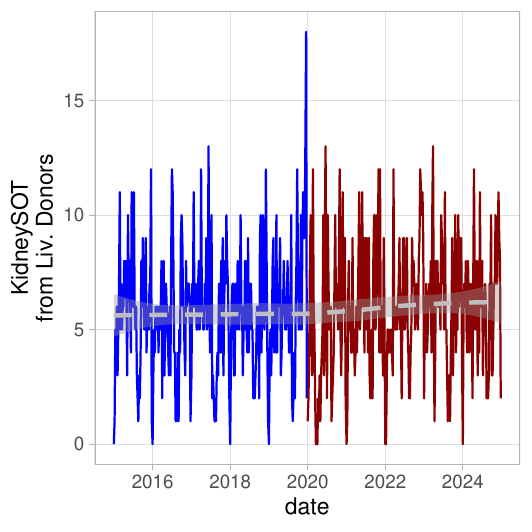}
        }

        \subfigure[Liver SOT from deceased donors.]{
            \includegraphics[width = 0.45\textwidth,height=0.25\textheight]{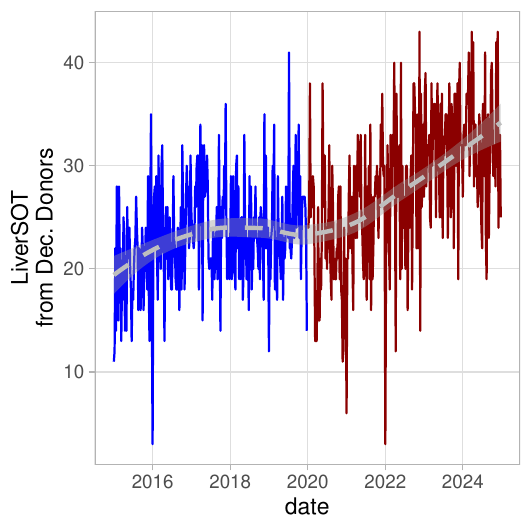}
        }
        \hfill
        \subfigure[Liver SOT from living donors.]{
            \includegraphics[width = 0.45\textwidth,height=0.25\textheight]{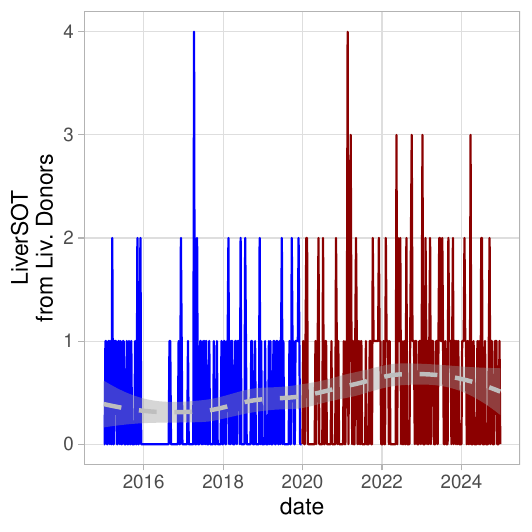}
        }
        \caption{Number of weekly transplants in Italy from deceased ((a),(c),(e)) and living ((b),(d),(f)) donors, comparing pre (blue line) and post (red line) COVID-19 time windows. Trends evaluated as loess curves are grey dashed lines.}
        \label{fig:dec-liv_IT}
    \end{figure}
   
    \begin{figure} 
        \centering
        \subfigure[deaths by all causes]{
        \includegraphics[width = 0.48\textwidth]{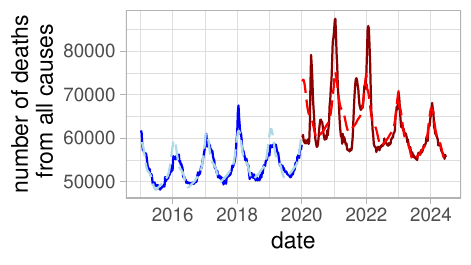}} 
       \subfigure[P-score]{ \includegraphics[width = 0.48\textwidth]{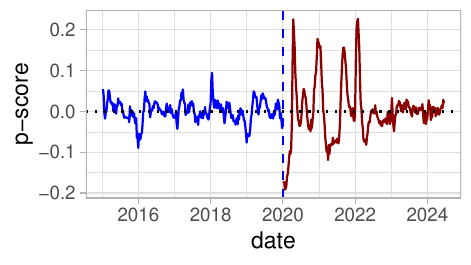}}
        \caption{USA; (a) Observed (continuous) and estimated (dashed) weekly deaths by all causes.  (b) Excess mortality from all causes evaluated as a p-score. Pre (blue) and post (red) COVID-19 time windows.}
        \label{fig:death+base_USA}
    \end{figure}
    \begin{figure} 
        \centering
        \subfigure[deaths by all causes]{
        \includegraphics[width = 0.48\textwidth]{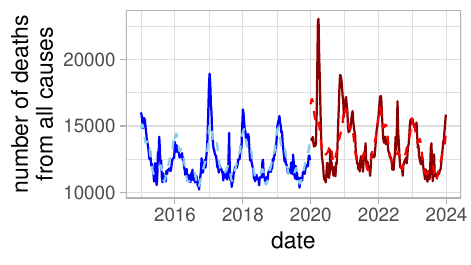}} 
       \subfigure[P-score]{ \includegraphics[width = 0.48\textwidth]{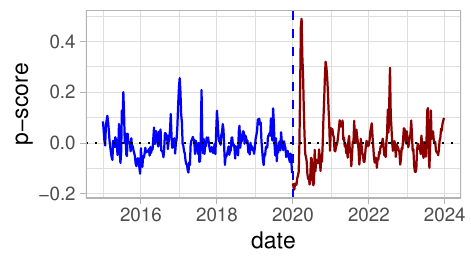}}
        \caption{Italy; (a) Observed (continuous) and estimated (dashed) weekly deaths by all causes.  (b) Excess mortality from all causes evaluated as a p-score. Pre (blue) and post (red) COVID-19 time windows.}
        \label{fig:death+base_IT}
    \end{figure}
    
    {In the USA, before the onset of the COVID-19 pandemic, a discernible upward trend is {apparent} upon analysing the number of transplants originating from {both} deceased and living donors  (Figure \ref{fig:dec-liv_USA} grey dashed lines). A notable decline in every time series is observed from 2020 to 2021. However, this significant reduction appears confined to a brief interval in 2020 (mostly between the week of March 25 and the week of May 6), coinciding with considerable strain on the healthcare system due to the unanticipated epidemic pressure. In the post-pandemic period, it seems that the increasing trend is again in place for transplants from deceased donors, {while living donors are starting at a lower level than pre-COVD-19. The yearly growth remains in place for all types of SOT. }
    In Italy, the temporal dynamics of transplant activity display a markedly different pattern compared to the USA (Figure \ref{fig:dec-liv_IT}, grey dashed lines). Before the COVID-19 outbreak, the number of transplants from deceased donors shows a moderate but steady upward trend. The pandemic caused an evident decline across all series between March and May 2020, mirroring the sharp disruption of routine hospital activities. In the post-pandemic period, the recovery from deceased donors gradually resume their pre-COVID trajectory, while living donations remain substantially stable over time.
    
    \par
    To depict the impact of the pandemic, we {start} by visually examining the resultant excess mortality. The right panels in Figures \ref{fig:death+base_USA} and \ref{fig:death+base_IT} illustrates the excess mortality quantified as the \textit{p-score} \citep{Pscore-excess-mortality}, which represents the proportion of the disparity between the reported and estimated death counts divided by the estimated number of deaths, a proven metric to assess the impact of the pandemic \citep{karlinsky2021tracking}. In both countries, the pandemic is visible as a shock in the p-score time series that is returning to a pre-COVID-19 condition starting from the end of 2022.  Italy exhibits markedly higher p-scores than the USA during both the first and second pandemic waves, followed by a final peak in the summer of 2022 attributable to the heatwave that affected the country. In contrast, the U.S. shows four distinct peaks of comparable magnitude over the same period.
    P{-}scores are evaluated from data in the left panels of  Figures \ref{fig:death+base_USA} and \ref{fig:death+base_IT}, where expected deaths were estimated following \cite{Maruottietal:2022}. The excess mortality can be seen as an indirect measure of the pressure on the country's health system. The preliminary findings support the idea of strong pressure on the entire health system in the US and Italy during the pandemic; however, despite this strain, Italy rapidly implemented safety protocols across healthcare settings, helping to largely contain the initial outbreak in the areas most affected and to limit wider spread.}
    

\subsection{Models estimation}
    To confirm the visual inspection as per Figures \ref{fig:dec-liv_USA} and \ref{fig:dec-liv_IT}, we searched for the best-fitting models for both countries and types of transplants. In addition, we performed the same model-selection procedure for both kidney and liver SOT from both deceased and living donors, except for liver transplants from living donors in Italy, for which the sample size was too limited to allow model estimation. \par
    The evaluation of the fitting ability of the best models selected are shown in Table \ref{tab:fittingRes} (USA) and Table \ref{tab:fittingRes_ITA} (Italy).
    All the chosen models can accurately capture the variability of the time series of SOT from both deceased and living transplants. Indeed, the empirical coverage is very close to the nominal value of 95\% for all the considered time series. In the USA, the ratio RMSE/Mean Count shows a stable $8\%$ error in SOT from deceased donors. Higher values are obtained for SOT from living donors (see Table \ref{tab:fittingRes}). The latter showing a more erratic behaviour than the SOT from deceased donors are subject to a larger fitting error. In Italy, similar patterns emerge but with higher relative errors. The RMSE-to-mean ratio is approximately 22\% for SOT from deceased donors. For living donors, the ratio rises to about 43–44\%, confirming a markedly higher variability and lower predictability of these time series. The higher relative errors observed in Italy can be largely attributed to the smaller number of transplants compared to the USA.
\begin{table}[H]
    \centering
    \begin{tabular}{|cc|cccc|}
      \hline
    Donors & Organ & Empirical  & RMSE & Mean  & RMSE/Mean \\ 
    &&Coverage&&Count&\\
      \hline
     & kidney & 0.943 & 27.12 & 329.8 & 0.082 \\ 
      Deceased & liver & 0.952 & 13.69 & 159.0 & 0.086 \\ 
     & total & 0.938 & 48.75 & 612.6 & 0.080 \\ \hline
       & kidney & 0.939 & 21.48 & 114.4 & 0.188 \\ 
      Living & liver & 0.959 & 3.39 & 9.0 & 0.376 \\ 
       & total & 0.945 & 22.84 & 123.5 & 0.185 \\ 
       \hline
    \end{tabular}
    \caption{USA models. Evaluation of fitting ability over the whole period (week 1 of 2014 to week 41 of 2024) for the best models selected according to the BIC.}
    \label{tab:fittingRes}
\end{table}

    \begin{table}[ht]
\centering
\begin{tabular}{|cc|cccc|}
  \hline
Donors & Organ & Empirical  & RMSE & Mean  & RMSE/Mean \\ 
&&Coverage&&Count&\\ 
  \hline
 & kidney & 0.97 & 7.99 & 34.00 & 0.23 \\ 
 Deceased  & liver & 0.96 & 5.61 & 25.40 & 0.22 \\ 
   & total & 0.96 & 14.95 & 67.00 & 0.22 \\ \hline
  Living & kidney & 0.96 & 2.56 & 5.80 & 0.44 \\ 
   & total & 0.94 & 2.72 & 6.30 & 0.43 \\ 
   \hline
  \end{tabular}
\caption{Italy models. Evaluation of fitting ability over the whole period (week 1 of 2014 to week 52 of 2024) for the best models selected according to the BIC.} 
\label{tab:fittingRes_ITA}
\end{table}
    
\subsubsection{Transplants from deceased donors}

\paragraph{USA}
    \begin{figure} 
        \centering
        \subfigure[]{
            \includegraphics[width = 0.45\textwidth,height=0.2\textheight]{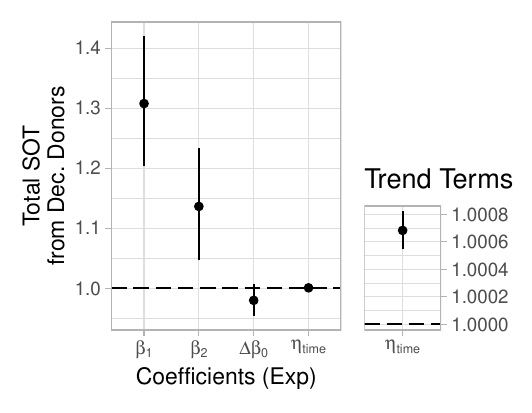}
        }
        \hfill
        \subfigure[]{
            \includegraphics[width = 0.45\textwidth,height=0.2\textheight]{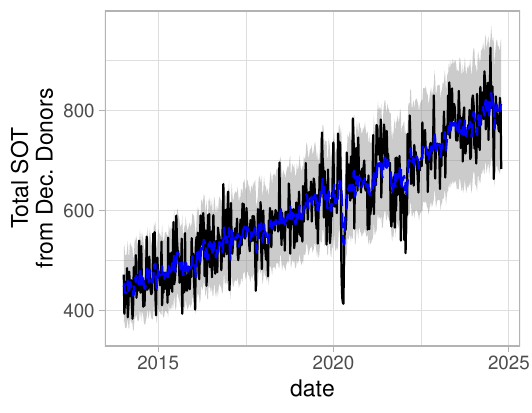}
        }
        \subfigure[]{
            \includegraphics[width = 0.45\textwidth,height=0.2\textheight]{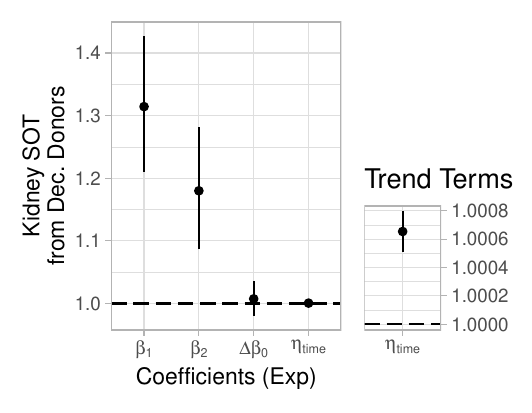}
        }
        \hfill
        \subfigure[]{
            \includegraphics[width = 0.45\textwidth,height=0.2\textheight]{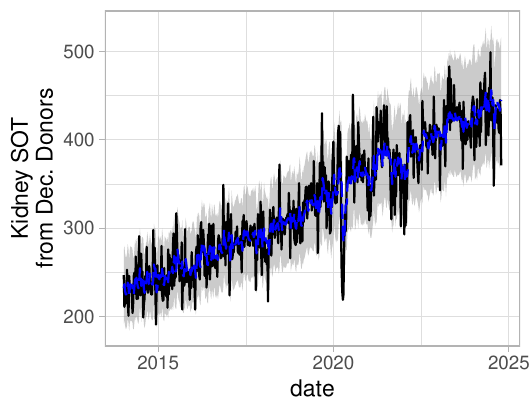}
        }

        \subfigure[]{
            \includegraphics[width = 0.45\textwidth,height=0.2\textheight]{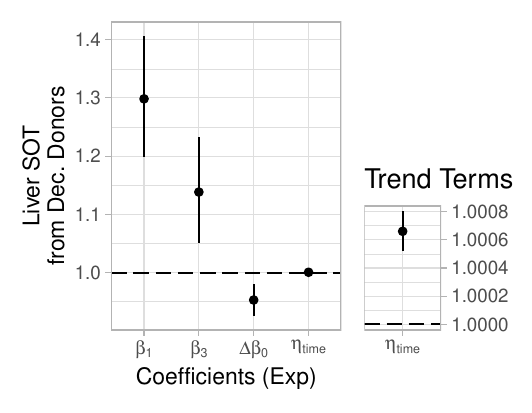}
        }
        \hfill
        \subfigure[]{
            \includegraphics[width = 0.45\textwidth,height=0.2\textheight]{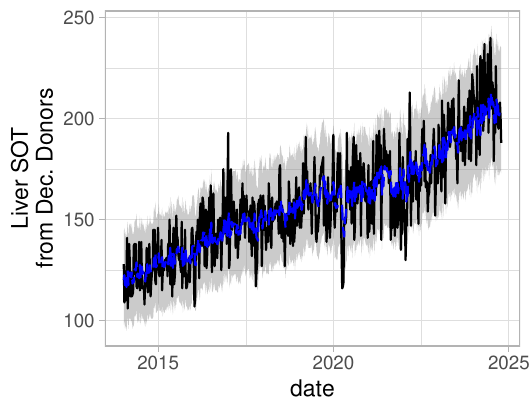}
        }
        \caption{USA's SOT from deceased donors: coefficients estimates on the exponential scale ((a),(c),(e)) with 95\% confidence limits,  fitted (blue) and observed (black) values  ((b),(d),(f)) with 95\% confidence band (grey) for the selected models.  (a)--(b) total number of transplants, (c)--(d) kidney transplants and (e)--(f) liver transplants}
        \label{fig:SOT_Deceased}
    \end{figure}

    According to the BIC, the selected model to describe total SOT from deceased donors is extremely parsimonious: it includes the intercept ($\beta_0$), short-term effects ($\beta_1,\beta_2$), a linear trend ($\eta_{time}$) and no significant variation between pre and post-COVID-19.  The full specification of the model and the parameter estimates are provided in the dedicated section of the Supplementary Materials (Table \ref{tab:dec}). This model suggests that the SOT from deceased donors did not mutate during the COVID-19 pandemic, as it excludes changes in both trend and intercept. \par
    
    Analysing the best (according to BIC) models for kidney and liver SOT from deceased donors, we observed that both of them share the same structure with the model for total transplants. The level of transplants, regardless of the type of organ, in each time-unit is positively influenced by the number of transplants recorded in the previous weeks ($\beta_1, \beta_2$, $\beta_3$), by the general mean effect in the same period and by a steady trend generating about a 4.4\% annual increase (computed as $\exp(52\cdot \eta_{time})$). Figure \ref{fig:SOT_Deceased} shows the models' coefficients on the exponential scale and fitted values together with 95\% confidence intervals. The full specification of the models and the parameter estimates are provided in the dedicated section of the Supplementary Materials (Section \ref{sec:sotsuplementary}).

\paragraph{Italy}
    \begin{figure} 
        \centering
        \subfigure[]{
            \includegraphics[width = 0.45\textwidth,height=0.2\textheight]{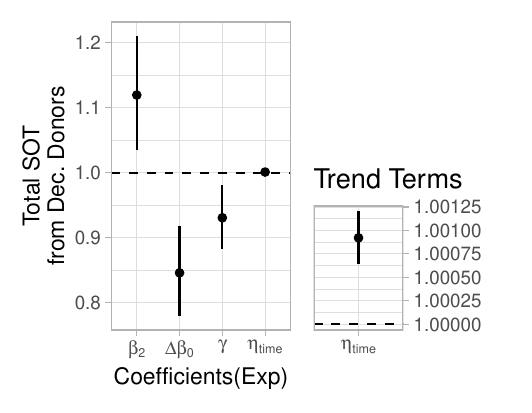}
        }
        \hfill
        \subfigure[]{
            \includegraphics[width = 0.45\textwidth,height=0.2\textheight]{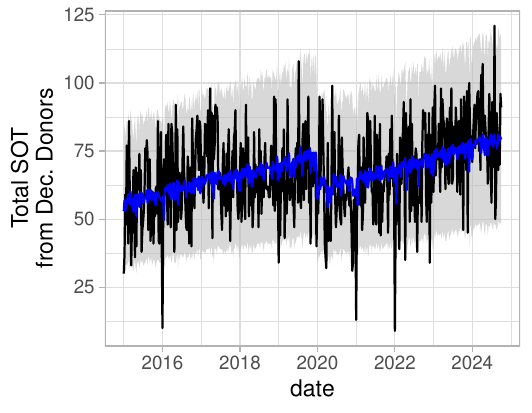}
        }
        \subfigure[]{ 
            \includegraphics[width = 0.45\textwidth,height=0.2\textheight]{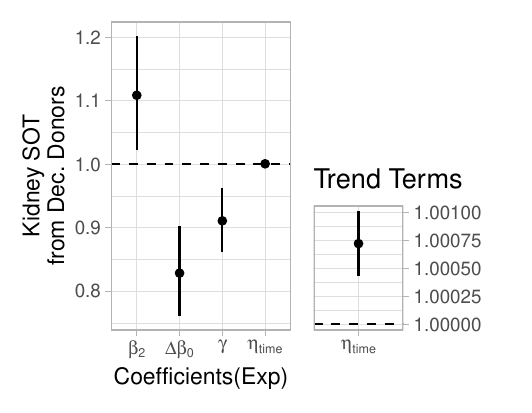}
        }
        \hfill
        \subfigure[]{
            \includegraphics[width = 0.45\textwidth,height=0.2\textheight]{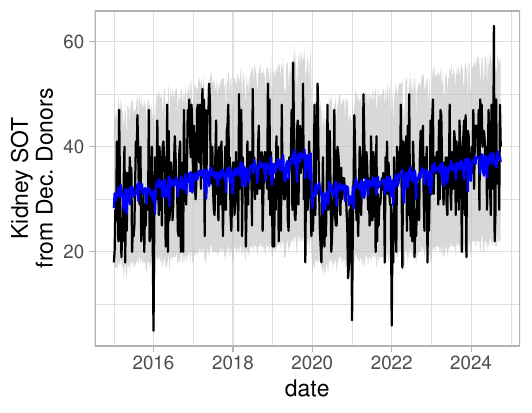}
        }

        \subfigure[]{
            \includegraphics[width = 0.45\textwidth,height=0.2\textheight]{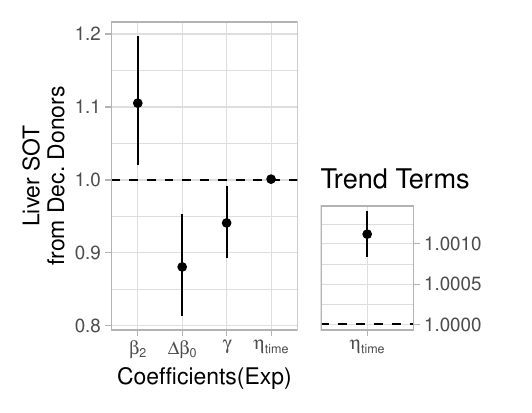}
        }
        \hfill
        \subfigure[]{
            \includegraphics[width = 0.45\textwidth,height=0.2\textheight]{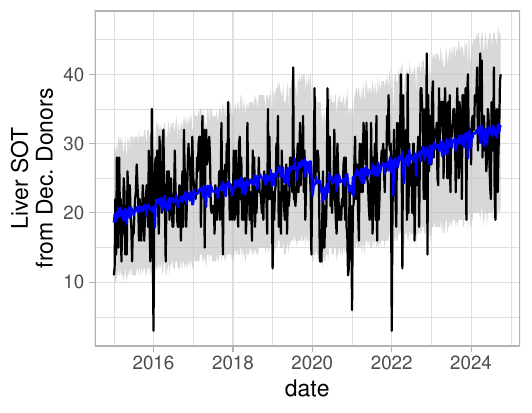}
        }
        \caption{Italy's SOT from deceased donors: coefficients estimates on the exponential scale ((a),(c),(e)) with 95\% confidence limits,  fitted (blue) and observed (black) values  ((b),(d),(f)) with 95\% confidence band (grey) for the selected models.  (a)--(b) total number of transplants, (c)--(d) kidney transplants and (e)--(f) liver transplants}
        \label{fig:SOT_Deceased_IT}
    \end{figure}
    According to the BIC, similarly to the US case, the selected models describing SOT from deceased donors in Italy are parsimonious. The model for total SOT from deceased donors includes the intercept ($\beta_0$), a short-term autoregressive component ($\beta_2$), a linear temporal trend ($\eta_{time}$), and, differently from the U.S., additional statistically-significant dummy covariates capturing the conditions before and after COVID-19 ($\Delta\beta_0$) and the effect of holiday weeks ($\gamma$). Specifically, we estimate a 14.0\% decrease in transplant activity during the COVID-19 period, a modest but significant 6.8\% reduction during holiday weeks, and a stable 4.3\% annual increase associated with the long-term trend.

The selected models (according to BIC) for kidney and liver SOT from deceased donors share the same general structure as the model for total transplants. Recent activity (captured by $\beta_2$) exerts a positive and significant influence, indicating short-term persistence in the number of transplants. The COVID-19 and holiday dummies remain significant and negative across organs, while the estimated weekly trends are positive and stable. Figure \ref{fig:SOT_Deceased_IT} displays the models’ coefficients on the exponential scale and the fitted values together with 95\% confidence intervals. The complete model specifications and parameter estimates are reported in the Supplementary Materials.
    
\subsubsection{Transplant from living donors}
\paragraph{USA}
    \begin{figure} 
        \centering
        \subfigure[]{
            \includegraphics[width = 0.45\textwidth,height=0.2\textheight]{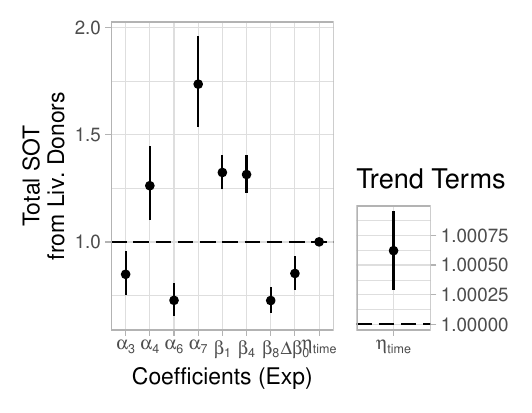}
        }
        \hfill
        \subfigure[]{
            \includegraphics[width = 0.45\textwidth,height=0.2\textheight]{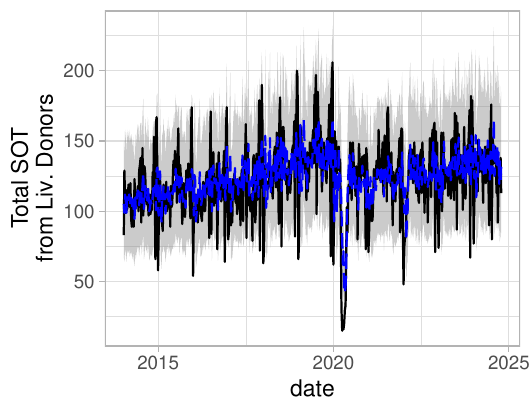}
        }
        \subfigure[]{
            \includegraphics[width = 0.45\textwidth,height=0.2\textheight]{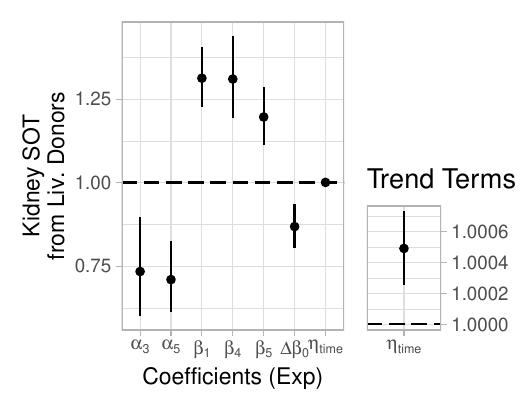}
        }
        \hfill
        \subfigure[]{
            \includegraphics[width = 0.45\textwidth,height=0.2\textheight]{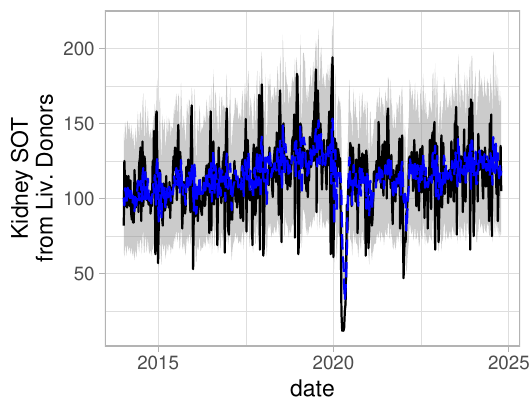}
        }
        
        \subfigure[]{
            \includegraphics[width = 0.45\textwidth,height=0.2\textheight]{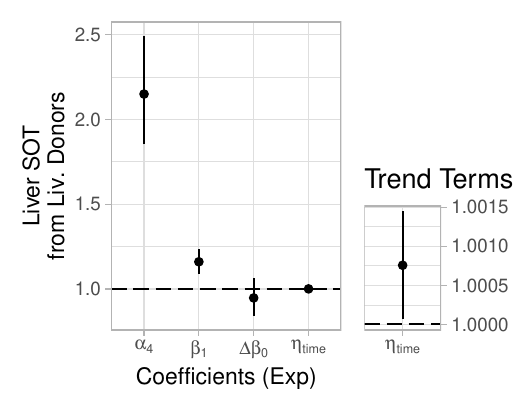}
        }
        \hfill
        \subfigure[]{
            \includegraphics[width = 0.45\textwidth,height=0.2\textheight]{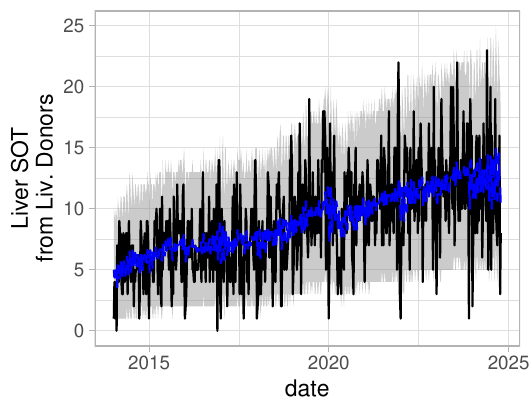}
        }
        \caption{USA's SOT from living donors: coefficients estimates on the exponential scale ((a),(c),(e)) with 95\% confidence limits,  fitted (blue) and observed (black) values  ((b),(d),(f)) with 95\% confidence band (grey) for the selected models.  (a)-(b) total number of transplants, (c)-(d) kidney transplants and (e)-(f) liver transplants}
        \label{fig:SOT_Living}
    \end{figure}
    The chosen model for Living donors reveals a {way} more {intricate} situation. The weekly number of SOT from living donors depends on a intercept ($\beta_0$), short-term effects up to one month before the date ($\beta_1, \beta_4, \alpha_3,\alpha_4$)  and long-term effects ($\beta_8,\alpha_6, \alpha_7$).  A linear trend ($\eta_{time}$) shows steady yearly growth in donors ($\exp(52\cdot 0.0006)$) of approximately 3\%. Again, using a dummy variable highlighting the conditions before and after COVID-19 ($\Delta\beta_0$) we see a 14.77\% reduction in the general level ($\beta_0$) on the exponential scale. Figure \ref{fig:SOT_Living} (a) and (b) show the model's coefficients on the exponential scale and fitted values together with 95\% confidence intervals.\par
    The full specification of the model and the parameter estimates are provided in the dedicated section of the Supplementary Materials (see also Table \ref{tab:liv} in the same section for point estimates of the parameters).\par
    The selected models for transplant from living donors by organ, kidney and liver include the intercept ($\beta_0$), and short-term effects ($\beta_1,\beta_4,\alpha_3$) for kidney, and ($\beta_1,\beta_2$) for liver. The kidney model includes long-term effects in both the mean and observation model parts ($\alpha_5$). Both models describe an increasing linear trend. For the kidney SOT from living donors, we still find a 13.3\% post-COVID-19 reduction on the exponential scale for the current level ($\Delta\beta_0$).\par
    Figure \ref{fig:SOT_Living} (c),(d) and (e),(f) shows the model's coefficients on the exponential scale and fitted values together with 95\% confidence intervals.\par
    The full specification of the model and the parameter estimates are provided in the dedicated section of the Supplementary Materials (see also Table \ref{tab:livKL} in the same section for point estimates of the parameters).\par
        
    Remarkably, once the time dynamic of the SOT is modelled, none of the selected models
    reports evidence of a statistically significant association between most considered factors. The only relevant effects are the trend (time) for both types of SOT and the COVID-19 effect still relevant for the living donors SOT.
    The latter supports the hypothesis of a full recovery of the health system for SOT from deceased donors, while a longer recovery time is required for the living donors.    \par

\paragraph{Italy}
\begin{figure} 
        \centering
        \subfigure[]{
            \includegraphics[width = 0.45\textwidth,height=0.2\textheight]{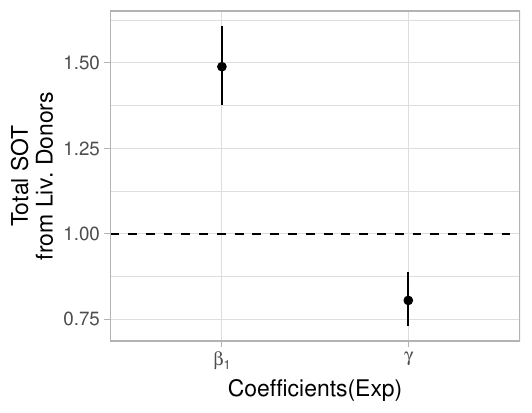}
        }
        \hfill
        \subfigure[]{
            \includegraphics[width = 0.45\textwidth,height=0.2\textheight]{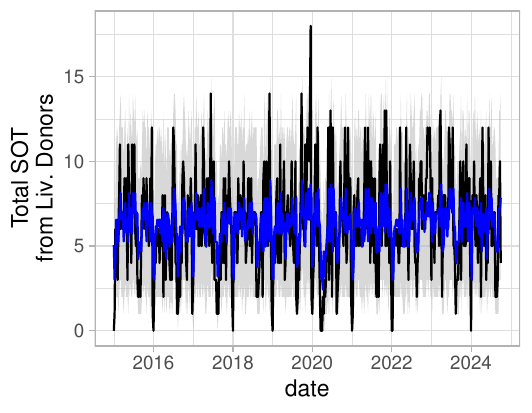}
        }
        \subfigure[]{
            \includegraphics[width = 0.45\textwidth,height=0.2\textheight]{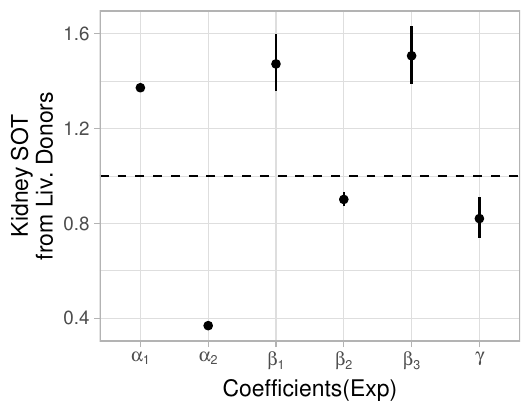}
        }
        \hfill
        \subfigure[]{
            \includegraphics[width = 0.45\textwidth,height=0.2\textheight]{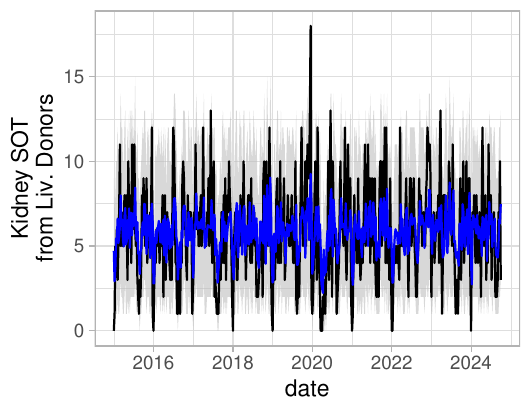}
        }
        
        \caption{Italy's SOT from living donors: coefficients estimates on the exponential scale ((a) and (c)) with 95\% confidence limits,  fitted (blue) and observed (black) values  ((b) and (d)) with 95\% confidence band (grey) for the selected models.  (a)-(b) total number of transplants and (c)-(d) kidney transplants}
        \label{fig:SOT_Living_IT}
    \end{figure}

    The models for SOT from living donors in Italy reveal a simpler temporal structure compared with those estimated for the USA. According to the BIC, the selected model for total SOT from living donors includes only the intercept ($\beta_0$), a short-term autoregressive component ($\beta_1$), and a dummy variable for holiday weeks ($\gamma$), all statistically significant. The negative coefficient for holidays indicates a 20.7\% reduction in transplant activity during those weeks, consistent with the well-known holiday-related decline driven by hospital organizational factors (reduced elective scheduling, limited staffing, and constrained operating room availability), and confirming the sensitivity of living donations to interruptions in routine hospital operations.

The selected model for kidney SOT from living donors follows a similar pattern but with a more complex autoregressive structure. It includes short-term dependencies ($\beta_1$, $\beta_3$ positive, and $\beta_2$ negative), and additional autoregressive effects in the conditional mean component ($\alpha_1$, $\alpha_2$). These terms capture a high short-term persistence in the number of living-donor kidney transplants, occasionally followed by compensatory fluctuations. The holiday dummy remains significant and negative, with an estimated 16.4\% reduction in transplant activity during holidays.

Unlike in the U.S., we did not find a \emph{robust} association with COVID-19-related indicators or mortality covariates once short-term dynamics and calendar effects were modelled; however, this statement is conditional on the explored model class and the (nested) selection protocol. Figure \ref{fig:SOT_Living_IT} shows the models’ coefficients on the exponential scale and fitted values together with 95\% confidence intervals. The full specification of the models and the parameter estimates are provided in the Supplementary Materials.

\subsection{Prediction performance}
We evaluate short- and medium-term forecasting performance using an expanding-window design described in Section~\ref{sec:forecast-eval}. For each outcome, we report point accuracy (RMSE) together with the empirical coverage of nominal $95\%$ predictive intervals. The corresponding summary tables for the USA and Italy are provided in the Supplementary Materials (Tables~\ref{tab:prediction} and \ref{tab:prediction_ITA}).

\section{Discussion}

Across outcomes and countries, the workflow consistently favored parsimonious count time-series specifications, with negative binomial models being selected when overdispersion was evident, and with short autoregressive structures. In addition, the COVID-19 intervention was captured through parametric terms driven by $C_t$ (level shift and/or slope change), while auxiliary covariates were never retained once the core dynamics and intervention terms were accounted for. This empirical selection pattern supports the use of simple, interpretable intervention-based count models to describe abrupt shocks in transplant activity.

The distributional predictive checks indicate that short- and medium-term forecasts are the most credible in this setting: longer-range forecasts would be dominated by the accumulation of uncertainty, and the selected models exhibit only short autoregressive dependence, suggesting limited predictability beyond a few weeks. 

Our results suggest a general strategy for biomedical count series subject to sudden shocks: (i) adopt a broad portfolio of parametric count time-series models augmented with pre-specified intervention terms capturing the shock; (ii) perform model selection within this portfolio using an information criterion; and (iii) validate forecasts using distributional metrics (empirical coverage of predictive intervals), focusing on short horizons when volatility is high. In the USA, although the cautious use of SARS-CoV-2 NAT+ non-lung donors has been reported as safe in terms of recipient outcomes \cite{martinez:2023}, our estimates suggest that living donation did not fully recover its pre-pandemic growth rate by 2023, with the reduction mainly observed in kidney transplants. Potential explanations include increased clinical caution regarding donor health, the reduced vaccine efficacy in SOT recipients, the persistent elevated risk of adverse outcomes following infection, and the psychological impact of the pandemic on health staff, patients and donors \cite{Nimmo:2022}. In Italy, where the first wave was earlier and more intense (Figure \ref{fig:death+base_IT}), the shock translated into pressure on intensive care units and a temporary reduction in deceased donor identification in 2020; however, the decline in transplant activity remained contained and temporally concentrated, with a rapid return toward the pre-pandemic trajectory. This is consistent with the early adoption of strict safety protocols by the Italian National Transplant Network (including systematic SARS-CoV-2 testing of deceased donors and COVID-free clinical pathways), and with the recognition of transplant activity as an essential level of care throughout the emergency.

Nevertheless, some limitations need to be noted. The workflow relies on large-scale model selection, and inference after selection should be interpreted with caution. The COVID-19 intervention is proxied by a pre-specified dummy at $t=(2020,1)$, which captures the main shock but cannot represent heterogeneous timing or multiple waves. The analysis is based on aggregated weekly counts, which may mask within-week variability and centre-level heterogeneity. Lastly, auxiliary covariates were considered within the portfolio, but unmeasured confounding and changes in clinical practice over time may still affect interpretation.

\subsection{Conclusions}
We presented a scalable, reproducible model-selection and validation workflow for modelling and forecasting weekly biomedical count time series under abrupt shocks, prioritizing distributional calibration (coverage) in addition to point accuracy. In the transplant case study (USA and Italy), the selected models provided interpretable estimates of disruption and recovery dynamics, indicating sustained regime differences in some strata and enabling operational summaries such as time-to-normality (defined in Methods) with uncertainty. The workflow is applicable to other healthcare activity series affected by shocks and reporting seasonality, and can be extended to multi-site hierarchical formulations and to explicit causal designs when suitable identification strategies are available.

\bibliographystyle{plainnat}
\bibliography{Bibliography.bib}

@article{ali2021safe,
  title={Is it safe to receive kidneys from deceased kidney donors tested positive for covid-19 infection?},
  author={Ali, Hatem and Mohamed, Mahmoud and Molnar, Miklos Z and Krishnan, Nithya},
  journal={Renal Failure},
  volume={43},
  number={1},
  pages={1060--1062},
  year={2021},
  publisher={Taylor \& Francis}
}

@article{gatti2022clinical,
  title={Clinical outcome in solid organ transplant recipients affected by {COVID-19} compared to general population: a systematic review and meta-analysis},
  author={Gatti, Milo and Rinaldi, Matteo and Bussini, Linda and Bonazzetti, Cecilia and Pascale, Renato and Pasquini, Zeno and Fan{\'\i}, Francesca and Guedes, Mariana Nunes Pinho and Azzini, Anna Maria and Carrara, Elena and others},
  journal={Clinical Microbiology and Infection},
  volume={28},
  number={8},
  pages={1057--1065},
  year={2022},
  publisher={Elsevier}
}

@article{vinson2021covid,
  title={{COVID-19} in solid organ transplantation: results of the national {COVID} cohort collaborative},
  author={Vinson, Amanda J and Agarwal, Gaurav and Dai, Ran and Anzalone, Alfred J and Lee, Stephen B and French, Evan and Olex, Amy and Madhira, Vithal and Mannon, Roslyn B},
  journal={Transplantation direct},
  volume={7},
  number={11},
  pages={e775},
  year={2021},
  publisher={LWW}
}

@article{coll2021covid,
  title={{COVID-19} in transplant recipients: the {Spanish} experience},
  author={Coll, Elisabeth and Fern{\'a}ndez-Ruiz, Mario and S{\'a}nchez-{\'A}lvarez, J Emilio and Mart{\'\i}nez-Fern{\'a}ndez, Jos{\'e} R and Crespo, Marta and Gayoso, Jorge and Bada-Bosch, Teresa and Oppenheimer, Federico and Moreso, Francesc and Lopez-Oliva, Maria O and others},
  journal={American Journal of Transplantation},
  volume={21},
  number={5},
  pages={1825--1837},
  year={2021},
  publisher={Elsevier}
}

@article{liboschik_tscount_2017,
	title = {tscount: An {R} Package for Analysis of Count Time Series Following {Generalized Linear Models}},
	volume = {82},
	rights = {Copyright (c) 2017 Tobias Liboschik, Konstantinos Fokianos, Roland Fried},
	issn = {1548-7660},
	url = {https://doi.org/10.18637/jss.v082.i05},
	doi = {10.18637/jss.v082.i05},
	shorttitle = {tscount},
	pages = {1--51},
	journal = {Journal of Statistical Software},
	author = {Liboschik, Tobias and Fokianos, Konstantinos and Fried, Roland},
	urldate = {2023-07-03},
	year = {2017},
    date = {2017-11-30},
	langid = {english},
}

@article{Porcuet:2022,
	author = {Hantouche, Mireille and Lara Carrion, Libia and Porcu, Emilio and Bramstedt, Katrina A.},
	date = {2022/11/30},
	doi = {10.1038/s41598-022-24351-x},
	id = {Hantouche2022},
	isbn = {2045-2322},
	journal = {Scientific Reports},
	number = {1},
	pages = {20651},
	title = {The effect of the {COVID}-19 pandemic on deceased and living organ donors in the United States of America},
	url = {https://doi.org/10.1038/s41598-022-24351-x},
	volume = {12},
	year = {2022}}

@article{Maruottietal:2022,
	author = {Maruotti, Antonello and Jona-Lasinio, Giovanna and Divino, Fabio and Lovison, Gianfranco and Ciccozzi, Massimo and Farcomeni, Alessio},
	date = {2022/02/01},
	doi = {10.1007/s40520-021-02060-1},
	id = {Maruotti2022},
	isbn = {1720-8319},
	journal = {Aging Clinical and Experimental Research},
	number = {2},
	pages = {475--479},
	title = {Estimating {COVID}-19-induced excess mortality in {Lombardy}, {Italy}},
	url = {https://doi.org/10.1007/s40520-021-02060-1},
	volume = {34},
	year = {2022}}

@article{Pscore-excess-mortality,
	author = {Janine Aron and John Muellbauer and Charlie Giattino and Hannah Ritchie},
	journal = {Our World in Data},
	note = {https://ourworldindata.org/covid-excess-mortality},
	title = {A pandemic primer on excess mortality statistics and their comparability across countries},
	year = {2020}}

@article{Nimmo:2022,
	author = {Nimmo, Ailish and Gardiner, Dale and Ushiro-Lumb, Ines and Ravanan, Rommel and Forsythe, John L. R.},
	id = {00007890-202207000-00010},
	isbn = {0041-1337},
	journal = {Transplantation},
	n2 = {The coronavirus disease 2019 (COVID-19) pandemic has had a major global impact on solid organ transplantation (SOT). An estimated 16{\%} global reduction in transplant activity occurred over the course of 2020, most markedly impacting kidney transplant and living donor programs, resulting in substantial knock-on effects for waitlisted patients. The increased severe acute respiratory syndrome coronavirus 2 (SARS-CoV-2) infection risk and excess deaths in transplant candidates has resulted in substantial effort to prioritize the safe restart and continuation of transplant programs over the second year of the pandemic, with transplant rates returning towards prepandemic levels. Over the past 2 y, COVID-19 mortality in SOT recipients has fallen from 20{\%}--25{\%} to 8{\%}--10{\%}, attributed to the increased and early availability of SARS-CoV-2 testing, adherence to nonpharmaceutical interventions, development of novel treatments, and vaccination. Despite these positive steps, transplant programs and SOT recipients continue to face challenges. Vaccine efficacy in SOT recipients is substantially lower than the general population and SOT recipients remain at an increased risk of adverse outcomes if they develop COVID-19. SOT recipients and transplant teams need to remain vigilant and ongoing adherence to nonpharmaceutical interventions appears essential. In this review, we summarize the global impact of COVID-19 on transplant activity, donor evaluation, and patient outcomes over the past 2 y, discuss the current strategies aimed at preventing and treating SARS-CoV-2 infection in SOT recipients, and based on lessons learnt from this pandemic, propose steps the transplant community could consider as preparation for future pandemics.},
	number = {7},
	title = {The Global Impact of {COVID}-19 on Solid Organ Transplantation: Two Years Into a Pandemic},
	url = {https://journals.lww.com/transplantjournal/fulltext/2022/07000/the_global_impact_of_covid_19_on_solid_organ.10.aspx},
	volume = {106},
	year = {2022}}

@article{AUBERT:2021,
title = {{COVID-19} pandemic and worldwide organ transplantation: a population-based study},
journal = {The Lancet Public Health},
volume = {6},
number = {10},
pages = {e709-e719},
year = {2021},
issn = {2468-2667},
doi = {https://doi.org/10.1016/S2468-2667(21)00200-0},
url = {https://www.sciencedirect.com/science/article/pii/S2468266721002000},
author = {Aubert, Olivier and Yoo, Daniel and Zielinski, Dina and Cozzi, Emanuele and Cardillo, Massimo and D{\"u}rr, Michael and Dom{\'\i}nguez-Gil, Beatriz and Coll, Elisabeth and Da Silva, Margarida Ivo and Sallinen, Ville and others}
}

@article{Ibrahim:2021,
	author = {Ibrahim, Ban and Dawson, Rosanne and Chandler, Jennifer A. and Goldberg, Aviva and Hartell, David and Hornby, Laura and Simpson, Christy and Weiss, Matthew-John and Wilson, Lindsay C. and Wilson, T. Murray and Fortin, Marie-Chantal},
	date = {2021/10/21},
	doi = {10.1186/s12910-021-00711-6},
	id = {Ibrahim2021},
	isbn = {1472-6939},
	journal = {BMC Medical Ethics},
	number = {1},
	pages = {142},
	title = {The {COVID}-19 pandemic and organ donation and transplantation: ethical issues},
	url = {https://doi.org/10.1186/s12910-021-00711-6},
	volume = {22},
	year = {2021}}

@article{SUAREZ:2022,
title = {Measuring the effect of the {COVID}-19 pandemic on solid organ transplantation},
journal = {The American Journal of Surgery},
volume = {224},
number = {1, Part B},
pages = {437-442},
year = {2022},
issn = {0002-9610},
doi = {https://doi.org/10.1016/j.amjsurg.2021.12.036},
url = {https://www.sciencedirect.com/science/article/pii/S0002961021007844},
author = {Alejandro Suarez-Pierre and Rashikh Choudhury and Adam M. Carroll and Robert W. King and John Iguidbashian and Jake Cotton and Kathryn L. Colborn and Peter T. Kennealey and Joseph C. Cleveland and Elizabeth Pomfret and David A. Fullerton}
}

@article{KUTE:2022,
title = {Global Impact of the {COVID}-19 Pandemic on {Solid Organ Transplant}},
journal = {Transplantation Proceedings},
volume = {54},
number = {6},
pages = {1412-1416},
year = {2022},
issn = {0041-1345},
doi = {https://doi.org/10.1016/j.transproceed.2022.02.009},
url = {https://www.sciencedirect.com/science/article/pii/S0041134522001324},
author = {Vivek B. Kute and Stefan G. Tullius and Hemant Rane and Sanshriti Chauhan and Vineet Mishra and Hari Shankar Meshram}
}

@article{FOKIANOS:2011,
title = {Log-linear {Poisson} autoregression},
journal = {Journal of Multivariate Analysis},
volume = {102},
number = {3},
pages = {563-578},
year = {2011},
issn = {0047-259X},
doi = {https://doi.org/10.1016/j.jmva.2010.11.002},
url = {https://www.sciencedirect.com/science/article/pii/S0047259X10002320},
author = {Konstantinos Fokianos and Dag Tj{\o}stheim}
}

@article{Chistou:2014,
author = {Christou, Vasiliki and Fokianos, Konstantinos},
title = {Quasi-likelihood inference for negative binomial time series models},
journal = {Journal of Time Series Analysis},
volume = {35},
number = {1},
pages = {55-78},
doi = {https://doi.org/10.1111/jtsa.12050},
url = {https://onlinelibrary.wiley.com/doi/abs/10.1111/jtsa.12050},
eprint = {https://onlinelibrary.wiley.com/doi/pdf/10.1111/jtsa.12050},
year = {2014}
}

@misc{coviddatawho,
	author = {WHO},
	lastchecked = {April,25,2024},
	title = {{COVID}-19 dashboard},
	urldate = {https://data.who.int/dashboards/covid19/data}
}

@misc{deathsdata,
	author = {OECD},
	lastchecked = {April,25,2024},
	title = {Data Explorer},
	urldate = {https://data-explorer.oecd.org/}
}

@book{timeSeries:2019,
	author = {Mills, Terence C.},
	publisher = {Academic Press, Elsevier},
	title = {Applied Time Series Analysis,  A Practical Guide to Modeling and Forecasting},
	year = {2019}}

@article{linearmixedModel:2021,
    author = {Verbeeck, Johan and Faes, Christel and Neyens, Thomas and Hens, Niel and Verbeke, Geert and Deboosere, Patrick and Molenberghs, Geert},
    title = "{A Linear Mixed Model to Estimate {COVID}-19-Induced Excess Mortality}",
    journal = {Biometrics},
    volume = {79},
    number = {1},
    pages = {417-425},
    year = {2021},
    month = {10},
    issn = {0006-341X},
    doi = {10.1111/biom.13578},
    url = {https://doi.org/10.1111/biom.13578},
    eprint = {https://academic.oup.com/biometrics/article-pdf/79/1/417/54629005/biometrics\_79\_1\_417.pdf},
}

@book{Hilbe_2014, place={Cambridge}, title={Modeling Count Data}, publisher={Cambridge University Press}, author={Hilbe, Joseph M.}, year={2014}}

@article{gneiting:2007,
author = {Tilmann Gneiting and Adrian E Raftery},
title = {Strictly Proper Scoring Rules, Prediction, and Estimation},
journal = {Journal of the American Statistical Association},
volume = {102},
number = {477},
pages = {359--378},
year = {2007},
publisher = {Taylor \& Francis},
doi = {10.1198/016214506000001437},


URL = { 
    
        https://doi.org/10.1198/016214506000001437
    
    

},
eprint = { 
    
        https://doi.org/10.1198/016214506000001437
    
    

}

}

@article{martinez:2023,
title = {Solid organ transplantation from donors with recent or current {SARS-CoV-2} infection: {A} systematic review},
journal = {Anaesthesia Critical Care \& Pain Medicine},
volume = {41},
number = {4},
pages = {101098},
year = {2022},
issn = {2352-5568},
doi = {https://doi.org/10.1016/j.accpm.2022.101098},
url = {https://www.sciencedirect.com/science/article/pii/S2352556822000790},
author = {Raquel Martinez-Reviejo and Sofia Tejada and Ana Cipriano and Hanife Nur Karakoc and Oriol Manuel and Jordi Rello}
}

@article{BIC,
author = {Gideon Schwarz},
title = {{Estimating the Dimension of a Model}},
volume = {6},
journal = {The Annals of Statistics},
number = {2},
publisher = {Institute of Mathematical Statistics},
pages = {461 -- 464},
year = {1978},
doi = {10.1214/aos/1176344136},
URL = {https://doi.org/10.1214/aos/1176344136}
}

@article{trapani2021incidence,
  title={Incidence and outcome of SARS-CoV-2 infection on solid organ transplantation recipients: a nationwide population-based study},
  author={Trapani, Silvia and Masiero, Lucia and Puoti, Francesca and Rota, Maria C and Del Manso, Martina and Lombardini, Letizia and Riccardo, Flavia and Amoroso, Antonio and Pezzotti, Patrizio and Grossi, Paolo A and others},
  journal={American Journal of Transplantation},
  volume={21},
  number={7},
  pages={2509--2521},
  year={2021},
  publisher={Elsevier}
}

@article{angelico2020covid,
  title={The COVID-19 outbreak in Italy: initial implications for organ transplantation programs},
  author={Angelico, Roberta and Trapani, Silvia and Manzia, Tommaso Maria and Lombardini, Letizia and Tisone, Giuseppe and Cardillo, Massimo},
  journal={American Journal of Transplantation},
  volume={20},
  number={7},
  pages={1780--1784},
  year={2020},
  publisher={Elsevier}
}

@article{karlinsky2021tracking,
  title={Tracking excess mortality across countries during the COVID-19 pandemic with the World Mortality Dataset},
  author={Karlinsky, Ariel and Kobak, Dmitry},
  journal={elife},
  volume={10},
  pages={e69336},
  year={2021},
  publisher={eLife Sciences Publications, Ltd}
}

@Manual{Rstudio,
    title = {R: A Language and Environment for Statistical Computing},
    author = {{R Core Team}},
    organization = {R Foundation for Statistical Computing},
    address = {Vienna, Austria},
    year = {2024},
    url = {https://www.R-project.org/},
  }

@article{schmidberger2009state,
  title={State of the Art in Parallel Computing with R},
  author={Schmidberger, Markus and Morgan, Martin and Eddelbuettel, Dirk and Yu, Hao and Tierney, Luke and Mansmann, Ulrich},
  journal={Journal of Statistical Software},
  volume={31},
  pages={1--27},
  year={2009}
}

@article{schwarz1978estimating,
  title={Estimating the dimension of a model},
  author={Schwarz, Gideon},
  journal={The annals of statistics},
  pages={461--464},
  year={1978},
  publisher={JSTOR}
}

\subsection*{List of abbreviations}
\begin{description}
  \item[BIC] Bayesian Information Criterion
  \item[CNT] National Transplant Centre (Centro Nazionale Trapianti)
  \item[ISS] Istituto Superiore di Sanit\`a
  \item[ISTAT] Italian National Institute of Statistics
  \item[NAT] Nucleic acid test
  \item[OECD] Organisation for Economic Co-operation and Development
  \item[RMSE] Root Mean Square Error
  \item[SOT] Solid organ transplant
\end{description}
\subsection*{Declarations}

\subsubsection*{Ethics approval and consent to participate}
Not Applicable

\subsubsection*{Consent for publication}
Not Applicable

\subsubsection*{Availability of data and materials}
Weekly transplant counts for the USA were obtained from publicly accessible sources, including the OECD data repository and the WHO data dashboard. Italian SOT's data are available from National Transplant Centre of the Istituto Superiore di Sanit\`a (ISS) but restrictions apply to the availability of these data, which were used under license for the current study, and so are not publicly available. Data are, however, available from the authors upon reasonable request and with permission of ISS. Requests for access to the SOT's dataset should be directed at: \href{mailto:silvia.trapani@iss.it}{silvia.trapani@iss.it}. Additional covariates were retrieved from publicly available sources, including ISTAT and the Italian Civil Protection.   

\subsubsection*{Competing interests}
The authors declare that they have no competing interests

\subsubsection*{Funding}
E.C. and G.J.L. were partially supported by the Sapienza University project ``Leaving No One Behind: Methods for Sustainability"  ID nr.: RD124190DA1146AA - CUP: B83C24007080005.

\subsubsection*{Authors' contributions}
G.J.L. and E.P. conceived the study. T.F. and E.C. wrote the code and performed the analysis. T.F., E.C., G.J.L. and E.P. wrote the article. E.D.S., S.P., S.Tr. and S.Te. provided the Italian SOT's data. E.P., L. L.-C., G.I., S.P, S.Tr. and S.Te. provided critical feedback. All the authors read and approved the final version of the paper.

\clearpage
\section*{Supplementary materials}

\subsection{Best models for SOT from deceased donors}

\paragraph{USA}
    \label{sec:sotsuplementary}
    The USA's models' specifications for total, kidney and liver SOT from deceased donors share the same structure. They are summarised in Equation (\ref{eq:bestDecModel}). Their relative estimates are provided in Tables \ref{tab:dec} and \ref{tab:decKL}. It is interesting to notice that the models' structures are extremely simple: they shows only lag-1 to lag-3 past observation terms. Furthermore, the value of $\eta_{time}=7\cdot 10^{-4}$ highlights the slow, yet steady, increase in the number of donors: it indicates that in 52 weeks, there is an increase of $\exp\round{52 \cdot 0.0007}=1.037$, that is a $3.7\%$ every year. Perhaps surprisingly, this number does not vary during the whole study period, as Figure \ref{fig:dec-liv_USA} clearly shows.
    \begin{align}
        \label{eq:bestDecModel}
        &Y_t \big| \mathcal{F}_{t-1} \sim \text{NegBin}\left(\exp(\mu_t + \eta_{time} t + \Delta\beta_0 \cdot C_t),\ \phi=\frac{1}{\sigma^2}\right)\\
        &\mu_t=\beta_0+\sum_{\ell=1}^3 \beta_\ell \log(Y_{t-\ell}+1)\nonumber
    \end{align}
    
    \begin{table}[H]
        \centering
        \begin{tabular}{lrrrr}
    \hline
    Model & Estimate & Std.Error & .95 CI   & .95 CI \\ 
    term &&&lower&upper\\
    \hline
$\beta_0$ & 3.7112 & 0.3010 & 3.1213 & 4.3012  \\ 
  $\beta_1$ & 0.2624 & 0.0420 & 0.1802 & 0.3447  \\ 
  $\beta_2$ & 0.1292 & 0.0421 & 0.0467 & 0.2118  \\ 
  $\eta_{time}$ & 0.0007 & 0.0001 & 0.0006 & 0.0008  \\ 
  $\Delta\beta_0$ & -0.0211 & 0.0138 & -0.0481 & 0.0058  \\ 
  $\sigma^2$ & 0.0050 &  &  &    \\ 
   \hline
\end{tabular}
        \caption{USA's estimated best model for total SOT from deceased donors. 
        95\%  confidence interval limits are reported. Link function: logarithm; distribution: negative binomial. BIC: 6014.7. Fitted coverage: 0.94.}
        \label{tab:dec}
    \end{table}
    
    \begin{table} [H]
        \centering

\begin{tabular}{lrrrr}
    \hline
    Model & Estimate & Std.Error & .95 CI   & .95 CI \\ 
    term &&&lower&upper\\
    \hline
    &&Kidney&&\\
    \hline
    $\beta_0$ & 3.0706 & 0.2648 & 2.5516 & 3.5895  \\ 
  $\beta_1$ & 0.2704 & 0.0418 & 0.1885 & 0.3524  \\ 
  $\beta_2$ & 0.1658 & 0.0420 & 0.0835 & 0.2481  \\ 
  $\eta_{time}$ & 0.0007 & 0.0001 & 0.0005 & 0.0008  \\ 
  $\Delta\beta_0$ & 0.0071 & 0.0140 & -0.0204 & 0.0346  \\ 
  $\sigma^2$ & 0.0040 &  &  &    \\ 
    \hline
     &&Liver&&\\
     \hline
    $\beta_0$ & 2.9229 & 0.2553 & 2.4224 & 3.4234  \\ 
  $\beta_1$ & 0.2588 & 0.0407 & 0.1789 & 0.3386  \\ 
  $\beta_3$ & 0.1306 & 0.0407 & 0.0508 & 0.2105  \\ 
  $\eta_{time}$ & 0.0007 & 0.0001 & 0.0005 & 0.0008 \\ 
  $\Delta\beta_0$ & -0.0489 & 0.0149 & -0.0781 & -0.0197  \\ 
  $\sigma^2$ & 0.0011 &  &  &    \\ 
    \hline
\end{tabular}
        \caption{USA's estimated best model for SOT from deceased donors by organ: kidney and liver. 
        95\%  confidence interval limits are reported. Link function: logarithm; distribution: negative binomial. BIC kidney model 5344.2, coverage: 0.94, BIC liver model 4562.2, coverage: 0.94.}
        \label{tab:decKL}
    \end{table}

    \paragraph{Italy}
    The Italy's models' specifications for total, kidney and liver SOT are identical, the modelling formula is reported in Equation \ref{eq:bestDecModel_ITA}. Their relative estimates are provided in Tables \ref{tab:dec_ITA} and \ref{tab:decKL_ITA}. The models structure is extremely simple, as it includes the intercept ($\beta_0$), a short-term effect ($\beta_2$), a linear trend ($\eta_{time}$), pre and post COVID-19 variation ($\Delta\beta_0$) and the holiday week effect ($\gamma$).

    \begin{align}
        \label{eq:bestDecModel_ITA}
        &Y_t \big| \mathcal{F}_{t-1} \sim \text{NegBin}\left(\exp(\mu_t + \eta_{time} t + \Delta\beta_0 \cdot C_t + \gamma\cdot\text{holiday}_t),\ \phi=\frac{1}{\sigma^2}\right)\\
        &\mu_t=\beta_0+\beta_2\cdot\log(Y_{t-2}+1)\nonumber
    \end{align}

    \begin{table} [H]
        \centering
        \begin{tabular}{lrrrr}
    \hline
    Model term & Estimate & Std.Error & .95 CI lower & .95 CI upper \\ 
    \hline
$\beta_0$ & 3.6065 & 0.1673 & 3.2785 & 3.9345 \\ 
$\beta_2$ & 0.1111 & 0.0412 & 0.0303 & 0.1919 \\ 
$\eta_{time}$ & 0.000803 & 0.000159 & 0.000491 & 0.001120 \\ 
$\Delta\beta_0$ & -0.1507 & 0.0433 & -0.2357 & -0.0658 \\ 
$\gamma$ & -0.0701 & 0.0286 & -0.1261 & -0.0141 \\ 
$\sigma^2$ & 0.0384 &  &  &  \\ 
\hline
\end{tabular}
        \caption{Italy's estimated best model for total SOT from deceased donors. 
        95\% confidence interval limits are reported. Link function: logarithm; distribution: negative binomial. BIC: 4272.0.}
        \label{tab:dec_ITA}
    \end{table}
    
    \begin{table} [H]
        \centering
        
\begin{tabular}{lrrrr}
    \hline
    Model term & Estimate & Std.Error & .95 CI lower & .95 CI upper \\ 
    \hline
    &&Kidney&&\\
    \hline
$\beta_0$         & 3.0871  & 0.1487 & 2.7956 & 3.3786 \\ 
$\beta_2$         & 0.1069  & 0.0428 & 0.0231 & 0.1907 \\ 
$\eta_{time}$     & 0.00059 & 0.00016 & 0.00027 & 0.00091 \\ 
$\Delta\beta_0$   & -0.1695 & 0.0449 & -0.2574 & -0.0816 \\ 
$\gamma$ & -0.0942 & 0.0298 & -0.1527 & -0.0357 \\ 
$\sigma^2$        & 0.0277  &         &         &        \\ 
    \hline
    &&Liver&&\\
    \hline
$\beta_0$         & 2.6815  & 0.1286 & 2.4294 & 2.9335 \\ 
$\beta_2$         & 0.0997  & 0.0422 & 0.0169 & 0.1824 \\ 
$\eta_{time}$     & 0.00116 & 0.00016 & 0.00084 & 0.00148 \\ 
$\Delta\beta_0$   & -0.1395 & 0.0419 & -0.2216 & -0.0574 \\ 
$\gamma$ & -0.0594 & 0.0283 & -0.1148 & -0.0040 \\ 
$\sigma^2$        & 0.0114  &         &         &        \\ 
    \hline
\end{tabular}
        \caption{Italy's estimated best model for SOT from deceased donors by organ: kidney and liver.
        95\%  confidence interval limits are reported. Link function: logarithm; distribution: negative binomial. BIC kidney model 3612.7; BIC liver model 3253.7.}
        \label{tab:decKL_ITA}
    \end{table}
    
\subsection{Best models for SOT from living donors}
\paragraph{USA}
    The models for the living donors SOT are slightly more complex than those in the previous section. The general model includes long and short-term effects (Equation \ref{eq:bestLivModel}),  mostly linked to the kidney SOT behaviour. The model's parameters include the intercept ($\beta_0$), short-term effects ($\beta_1, \beta_4, \alpha_3,\alpha_4$) and  long-term effects ($\beta_8,\alpha_6, \alpha_7$) a linear trend ($\eta_{time}$) and a dummy variable to highlight pre and post COVID-19 conditions ($\Delta\beta_0$) highliting a 14.77\% reduction in the intercept on the exponential scale. Yearly growth $\exp(52\cdot 0.0006)$ indicates a 3\% steady increase in donors.
    
    Indeed, liver SOT shows a much simpler structure, similar to the SOT from deceased donors. Both kidney and liver models include the intercept ($\beta_0$), and short-term effects ($\beta_1,\beta_4,\alpha_3$ for kidney, $\beta_1,\beta_2$ for liver). The kidney model includes long-term effects in both the mean and observation model parts ($\alpha_5$). Both models describe an increasing linear trend. For the kidney SOT from living donors, we still find a 13.3\% reduction on the exponential scale for the intercept.
    \begin{align}
        \label{eq:bestLivModel}
        &Y_t \big| \mathcal{F}_{t-1} \sim \text{NegBin}\left(\exp(\mu_t + \eta_{time}t + \Delta\beta_0 \cdot C_t), \phi=\frac{1}{\sigma^2}\right)\\    
        &\mu_t=\beta_0 + \sum_{\ell=1}^8\beta_\ell \log(Y_{t-\ell}+1) +\sum_{\ell=1}^4 \alpha_\ell \mu_{t-\ell} \nonumber
    \end{align}
    \begin{table} [H]
        \centering
        \begin{tabular}{lrrrr}
    \hline
    Model & Estimate & Std.Error & .95 CI   & .95 CI \\ 
    term &&&lower&upper\\
    \hline
    $\beta_0$ & 2.0918 & 0.5144 & 1.0836 & 3.1000  \\ 
  $\beta_1$ & 0.2797 & 0.0309 & 0.2191 & 0.3403  \\ 
  $\beta_4$ & 0.2729 & 0.0334 & 0.2075 & 0.3383  \\ 
  $\beta_8$ & -0.3255 & 0.0409 & -0.4056 & -0.2454  \\ 
  $\alpha_3$ & -0.1640 & 0.0614 & -0.2843 & -0.0436  \\ 
  $\alpha_4$ & 0.2351 & 0.0681 & 0.1017 & 0.3685  \\ 
  $\alpha_6$ & -0.3107 & 0.0532 & -0.4150 & -0.2065  \\ 
  $\alpha_7$ & 0.5632 & 0.0602 & 0.4452 & 0.6812  \\ 
  $\eta_{time}$ & 0.0006 & 0.0002 & 0.0003 & 0.0010  \\ 
  $\Delta\beta_0$ & -0.1599 & 0.0477 & -0.2534 & -0.0664  \\ 
  $\sigma^2$ & 0.0287 &  &  &    \\ 
   \hline
\end{tabular}
        \caption{USA's estimated best model for total SOT from living donors. 
        95\%  confidence interval limits are reported. Link function: logarithm, likelihood: negative binomial. BIC: 5271.9.}
        \label{tab:liv}
    \end{table}

 \begin{table} [H]
        \centering
        
\begin{tabular}{lrrrr}
    \hline
    Model & Estimate & Std.Error & .95 CI   & .95 CI \\ 
    term &&&lower&upper\\
    \hline
    &&Kidney&&\\
    \hline
$\beta_0$ & 4.3001 & 0.5038 & 3.3126 & 5.2877  \\ 
  $\beta_1$ & 0.2716 & 0.0349 & 0.2031 & 0.3400  \\ 
  $\beta_4$ & 0.2709 & 0.0478 & 0.1772 & 0.3647  \\ 
  $\beta_5$ & 0.1776 & 0.0370 & 0.1052 & 0.2501  \\ 
  $\alpha_3$ & -0.3108 & 0.1023 & -0.5113 & -0.1102  \\ 
  $\alpha_5$ & -0.3428 & 0.0762 & -0.4922 & -0.1934  \\ 
  $\eta_{time}$ & 0.0005 & 0.0001 & 0.0003 & 0.0007  \\ 
  $\Delta\beta_0$ & -0.1425 & 0.0386 & -0.2181 & -0.0668  \\ 
  $\sigma^2$ & 0.0293 &  &  &    \\ 

\hline
 &&Liver&&\\
     \hline
$\beta_0$ & 3.7112 & 0.3010 & 3.1213 & 4.3012  \\ 
  $\beta_1$ & 0.2624 & 0.0420 & 0.1802 & 0.3447  \\ 
  $\beta_2$ & 0.1292 & 0.0421 & 0.0467 & 0.2118  \\ 
  $\eta_{time}$ & 0.0007 & 0.0001 & 0.0006 & 0.0008  \\ 
  $\Delta\beta_0$ & -0.0211 & 0.0138 & -0.0481 & 0.0058  \\ 
  $\sigma^2$ & 0.0050 &  &  &   \\ 
\hline
\end{tabular}

        \caption{USA's estimated best model for SOT from living donors by organ: kidney and liver. 
        95\% confidence limits are reported. BIC kidney model 5201.7, BIC liver model 2994. } 
        \label{tab:livKL}
    \end{table}

\paragraph{Italy}

The model for total SOT from living donors in Italy displays a much simpler temporal structure than those estimated for the United States. According to the BIC, the best model for total SOT from living donors includes only the intercept ($\beta_0$), a short-term autoregressive component ($\beta_1$) and a dummy variable for holiday weeks ($\gamma$). The negative holiday effect corresponds to an average reduction of about 19\% in transplant activity during those weeks. We did not find a stable trend or COVID-period effect once short-term dependence and holiday weeks were accounted for; this statement is conditional on the explored model class and selection protocol. The full model specification is reported in Equation \ref{eq:bestLivModel_IT} and parameter estimates are reported in Table~\ref{tab:livITA}.

For kidney SOT from living donors, the best model (according to BIC) retains a richer autoregressive structure. In addition to the intercept ($\beta_0$), it includes short-term dependencies ($\beta_1, \beta_2$ and $\beta_3$), and autoregressive terms in the mean component ($\alpha_1, \alpha_2$). The holiday dummy's coefficient $\gamma$ remains significant and negative, with an average reduction of about 15\% in transplant numbers during holidays. As for total living donations, no stable trend or COVID-period effect is identified under the selected specification. The complete model specifications and parameter estimates are reported in Table~\ref{tab:livKITA}. No model was estimated for liver SOT from living donors in Italy due to the very limited number of observations.

\begin{align}
        \label{eq:bestLivModel_IT}
        &Y_t \big| \mathcal{F}_{t-1} \sim \text{Poi}\left(\exp(\mu_t + \gamma\cdot \text{holiday}_t)\right)\\    
        &\mu_t=\beta_0 + \beta_1\cdot \log(Y_{t-1}+1) \nonumber \nonumber
    \end{align}
\begin{table}[H]
  \centering
  \begin{tabular}{lrrrr}
    \hline
    Model term & Estimate & Std.Error & .95 CI lower & .95 CI upper \\ 
    \hline
$\beta_0$ & 1.0460 & 0.0866 & 0.8760 & 1.2160 \\ 
$\beta_1$             & 0.4240 & 0.0428 & 0.3400 & 0.5080 \\ 
$\gamma$    & -0.2070 & 0.0542 & -0.3130 & -0.1010 \\ 
    \hline
\end{tabular}
  \caption{Italy's estimated best model for total SOT from living donors. 95\% confidence interval limits are reported. BIC 2474; Link function: logarithm; likelihood: Poisson.}
  \label{tab:livITA}
\end{table}

\begin{table}[H]
  \centering
  \begin{tabular}{lrrrr}
    \hline
    Model term & Estimate & Std.Error & .95 CI lower & .95 CI upper \\ 
    \hline
    &&Kidney&&\\
    \hline
$\beta_0$ & 1.6540 & 0.1496 & 1.3600 & 1.9467 \\ 
$\beta_1$             & 0.4390 & 0.0449 & 0.3520 & 0.5274 \\ 
$\beta_2$             & -0.1210 & 0.0169 & -0.1540 & -0.0875 \\ 
$\beta_3$             & 0.4470 & 0.0442 & 0.3600 & 0.5333 \\ 
$\alpha_1$            & 0.2580 & 0.0172 & 0.2240 & 0.2913 \\ 
$\alpha_2$            & -1.0000 & 0.0118 & -1.0230 & -0.9769 \\ 
$\gamma$    & -0.1640 & 0.0579 & -0.2770 & -0.0502 \\ 
    \hline
\end{tabular}
  \caption{Italy's estimated best model for kidney SOT from living donors in Italy. 95\% confidence interval limits are reported. BIC 2428.042; Link function: logarithm; likelihood: Poisson.}
  \label{tab:livKITA}
\end{table}
\section{Forecasts}
    \begin{table} [H]
        \centering
       
\begin{tabular}{llrr}
  \hline
Model & Weeks & RMSE & Coverage \\ 
  \hline
  Deceased Donors&&&\\\hline
 &  4 & 38.8 & 0.75 \\ 
  Kidney SOT  &  8 & 38.0 & 0.88 \\ 
 & 12 & 31.6 & 0.92 \\ \hline
   &  4 & 12.3 & 1.00 \\ 
  Liver SOT  &  8 & 9.8 & 1.00 \\ 
   & 12 & 11.9 & 1.00 \\ \hline
   &  4 & 66.9 & 0.75 \\ 
  Total SOT  &  8 & 56.8 & 0.88 \\ 
  & 12 & 49.8 & 0.92 \\ \hline
   Living Donors&&&\\\hline
   &  4 & 15.6 & 1.00 \\ 
  Kidney SOT  &  8 & 21.1 & 1.00 \\ 
   & 12 & 18.6 & 1.00 \\ \hline
   &  4 & 5.3 & 0.75 \\ 
  Liver SOT  &  8 & 5.0 & 0.88 \\ 
  & 12 & 4.6 & 0.92 \\ \hline
  &  4 & 16.7 & 1.00 \\ 
  Total SOT  &  8 & 22.9 & 1.00 \\ 
   & 12 & 20.9 & 1.00 \\ 
   \hline
\end{tabular}

        \caption{Evaluation of the predictive capacity of the chosen models in the USA. RMSE and empirical coverage of nominal $95\%$ predictive intervals are reported for different time horizons ($h=4,8,12$ weeks). }
        \label{tab:prediction}
    \end{table}
    
    \begin{table} [H]
        \centering
       
\begin{tabular}{llrr}
  \hline
Model & Weeks & RMSE & Coverage \\ 
  \hline
  Deceased Donors&&&\\\hline
   &  4 & 7.02 & 1.00 \\ 
  Kidney SOT  &  8 & 10.21 & 1.00 \\ 
   & 12 & 10.29 & 0.92 \\ \hline
   &  4 & 2.29 & 1.00 \\ 
  Liver SOT  &  8 & 5.90 & 1.00 \\ 
   & 12 & 5.58 & 1.00 \\ \hline
   &  4 & 9.15 & 1.00 \\ 
  Total SOT  &  8 & 18.99 & 1.00 \\ 
   & 12 & 18.54 & 1.00 \\ \hline
  Living Donors&&&\\\hline
   &  4 & 1.50 & 1.00 \\ 
  Kidney SOT  &  8 & 2.24 & 1.00 \\ 
   & 12 & 3.33 & 0.92 \\ \hline
   &  4 & 1.37 & 1.00 \\ 
  Total SOT  &  8 & 2.15 & 1.00 \\ 
   & 12 & 2.76 & 0.92 \\ 
  \hline
\end{tabular}

        \caption{Evaluation of the predictive capacity of the chosen models in Italy. RMSE and empirical coverage of nominal $95\%$ predictive intervals are reported for different time horizons ($h=4,8,12$ weeks).}
        \label{tab:prediction_ITA}
    \end{table}

\end{document}